 \documentclass[final,3p,times,twocolumn]{elsarticle}

\usepackage{amssymb}

\usepackage{graphicx, subfigure}
\usepackage{placeins}
\usepackage{float}

\newcommand{\footlabel}[2]{%
\addtocounter{footnote}{1}%
\footnotetext[\thefootnote]{%
\addtocounter{footnote}{-1}%
\refstepcounter{footnote}\label{#1}%
#2%
}%
$^{\ref{#1}}$%
}

\newcommand{\footref}[1]{%
$^{\ref{#1}}$%
}

\usepackage[%
backref=true,%
pagebackref=true,%
hyperfootnotes=true,%
hyperindex=true,%
pageanchor=true,%
plainpages=false,%
colorlinks=true,%
linkcolor=blue,%
citecolor=red,%
urlcolor=cyan,%
linktocpage=true,%
pdfstartview={Fit},
bookmarksopen=false
]{hyperref}
\usepackage{color}
\usepackage{hyperref}

\journal{Astroparticle Physics}

\begin{document}

\begin{frontmatter}


\title{On the potential of atmospheric Cherenkov telescope arrays for resolving TeV gamma-ray sources in the Galactic Plane}
 \author[gssi]{L. Ambrogi}  \ead{lucia.ambrogi@gssi.infn.it}
  \author[csic]{E. De O\~na Wilhelmi}
 \author[gssi,dias,mpik]{F. Aharonian}
 \address[gssi]{Gran Sasso Science Institute, viale Francesco Crispi, 7 67100 L'Aquila (AQ), Italy} 
 \address[csic]{Institute for Space Sciences (CSIC/IEEC), E-08193 Barcelona, Spain}
 \address[dias]{Dublin Institute of Advanced Studies, 10 Burlington Road, Dublin 4, Ireland} 
 \address[mpik]{Max-Planck-Institut f\"ur Kernphysik, Saupfercheckweg 1, D-69117 Heidelberg, Germany} 


\begin{abstract}
The potential of an array of imaging atmospheric Cherenkov telescopes to detect gamma-ray sources in complex regions has been investigated. The basic characteristics of the gamma-ray instrument have been parametrized using simple analytic representations. In addition to the ideal (Gaussian form) point spread function (PSF), the impact of more realistic non-Gaussian PSFs with tails has been considered. Simulations of isolated point-like and extended sources have been used as a benchmark to test and understand the response of the instrument. The capability of the instrument to resolve multiple sources has been analyzed and the corresponding instrument sensitivities calculated. The results are of particular interest for weak gamma-ray emitters located in crowded regions of the Galactic plane, where the chance of clustering of two or more gamma-ray sources within 1 degree is high.
\end{abstract}


\end{frontmatter}


\section{Introduction} 
\label{introduction}

High energy gamma rays are ideal carriers of information about non thermal relativistic processes in astrophysical objects. They are copiously produced in many Galactic and Extragalactic sources and freely propagate in space without deflection by interstellar and intergalactic magnetic fields. At very high energies, gamma-rays can be effectively detected by ground-based instruments. Among various techniques, the Imaging Atmospheric Cherenkov Telescope (IACT) technique has proven to be the most sensitive approach \citep{0034-4885-71-9-096901, doi:10.1146/annurev.nucl.47.1.273, Hinton:2009zz, Hillas:2013txa}.

In recent years, ground based telescopes such as H.E.S.S. \cite{1742-6596-60-1-021}, MAGIC \cite{2008JPhCS.110f2009F} and VERITAS \cite{Holder:2008ux}, have demonstrated the great potential offered by IACT stereoscopic arrays. These instruments have proven themselves as effective multifunctional tools for spectral, temporal, and morphological studies of gamma-ray sources at energies above a few tens of GeV.

The optimal instrument configuration for a next-generation IACT array should follow from the particular science goals. For instance, the studies of blazars and Gamma-Ray Bursts demand very low energy thresholds to extend the gamma-ray horizon to cosmological distances. On the other hand, for TeV Galactic sources, e.g. Supernova Remnants and Pulsar Wind Nebulae, the extension of the energy coverage up to 100 TeV is of prime importance. 

In the \emph{sub-TeV regime} the main target is to lower the energy threshold. Lower energy gamma-rays produce smaller showers and correspondingly less Cherenkov light. The energy threshold can be lowered by increasing the mirror area of the telescopes, by using focal plane detectors with high quantum efficiency or by placing the instrument at higher altitude (see e.g. ref. \cite{Aharonian:2000rf}).

In the \emph{TeV regime}, the flux level sharply drops due to the typical power-law spectral shape of non-thermal processes. Thus, it is necessary to maximize the effective area in this energy domain. 
This can be achieved with a large array of wide-field telescopes spread-out in a grid covering an area of several square kilometers.

The upcoming next-generation IACT array, the Cherenkov Telescope Array (CTA) \citep{CTA}, will exploit three different sizes of telescopes, each one optimized for low (multi-GeV), medium (TeV) and high (multi-TeV) energies. Thanks to its wide energy range, excellent angular and energy resolution and huge detection area, CTA is expected to provide a deep insight into the non-thermal Universe. 

In this paper, we use the CTA observatory as a template for the description of a generic IACT array. With simple analysis methods we address questions of primary importance for planning IACTs observations. We demonstrate the importance of such questions for drawing general conclusions independent of the specific instrument details, since at this stage absolute results might not be meaningful given the possible changes in the final layout design of CTA. With these ultimate goals, we study the potential of the array for the detection of weak gamma-ray sources in complex environments in the presence of multiple strong TeV emitters.  
 
\section{Instrument response} 
\label{cap:detector}

To investigate the performance of the telescope array, the simulated sources need to be convolved with the instrument response. Given the different performance at different energy intervals and, in general, different physics goals, we distinguish four energy intervals: [0.05-0.1]\,TeV, [0.1-1]\,TeV, [1-10]\,TeV and [10-100]\,TeV. 
This is the wide energy region covered by the CTA South array, particularly suitable for observation of the Galactic plane which is rich in TeV emitters belonging to several source populations. A total of 4 large-size, a few tens of medium-size and about 70 small-size telescopes will form the array at the CTA southern site, whose layout is under consideration. The instrument response functions for the southern array of CTA we used in this work consist of 4 large-size telescopes, 24 medium-size telescopes and 72 small-size telescopes. The performance is obtained by Monte Carlo (MC) simulations of a point-like gamma-ray source with a spectral shape similar to that of the Crab Nebula, located at the centre of the field of view (FoV) and observed at a zenith angle of 20 degrees \citep{Hassan:2015bwa}. From these publicly available distributions\footlabel{CTAwebpage}{The CTA performance files can be accessed at \url{https://portal.cta-observatory.org/Pages/CTA-Performance.aspx}. } 
simple analytical parameterizations in the energy range from 50\,GeV to 100\,TeV are derived for the angular resolution, effective area, energy resolution and background rate per unit of solid angle after rejection cuts. 
In table \ref{tab:response} the values for the angular resolution, effective area and background rate are shown for all four energy intervals.

\begin{table}
\begin{center}
\resizebox{\linewidth}{!}{
\begin{tabular}{ | c | c | c | c | }
  \hline                       
  Energy & $\sigma_{PSF}$ $[\mbox{deg}]$ & $A_{eff}$ $[\mbox{m}^2]$ &  $BgRate$ $[\mbox{Hz/deg}^2]$ \\
  \hline
   $[0.05-0.1$]\,TeV	& $0.147$	&  $4.1\cdot10^4$ 	&  $9.69\cdot10^{-1}$ \\
   $[0.1-1]$\,TeV	& $0.083$ &  $2.4\cdot10^5$	&  $1.53\cdot10^{-1}$\\
   $[1-10]$\,TeV		& $0.042$	&  $1.66\cdot10^6$	&  $3.20\cdot10^{-3}$	\\
   $[10-100]$\,TeV	& $0.031$ & $3.73\cdot10^6$	&  $3.55\cdot10^{-5}$ \\
  \hline 
\end{tabular}
}
\caption{\label{tab:response}The angular resolution, effective area and background rate per square degree in the four energy intervals.}
\end{center}
\end{table}

\subsection{Angular resolution} 
\label{cap:angres}
\begin{figure}
\centering
\includegraphics[width=8.5cm,height=5.8cm]{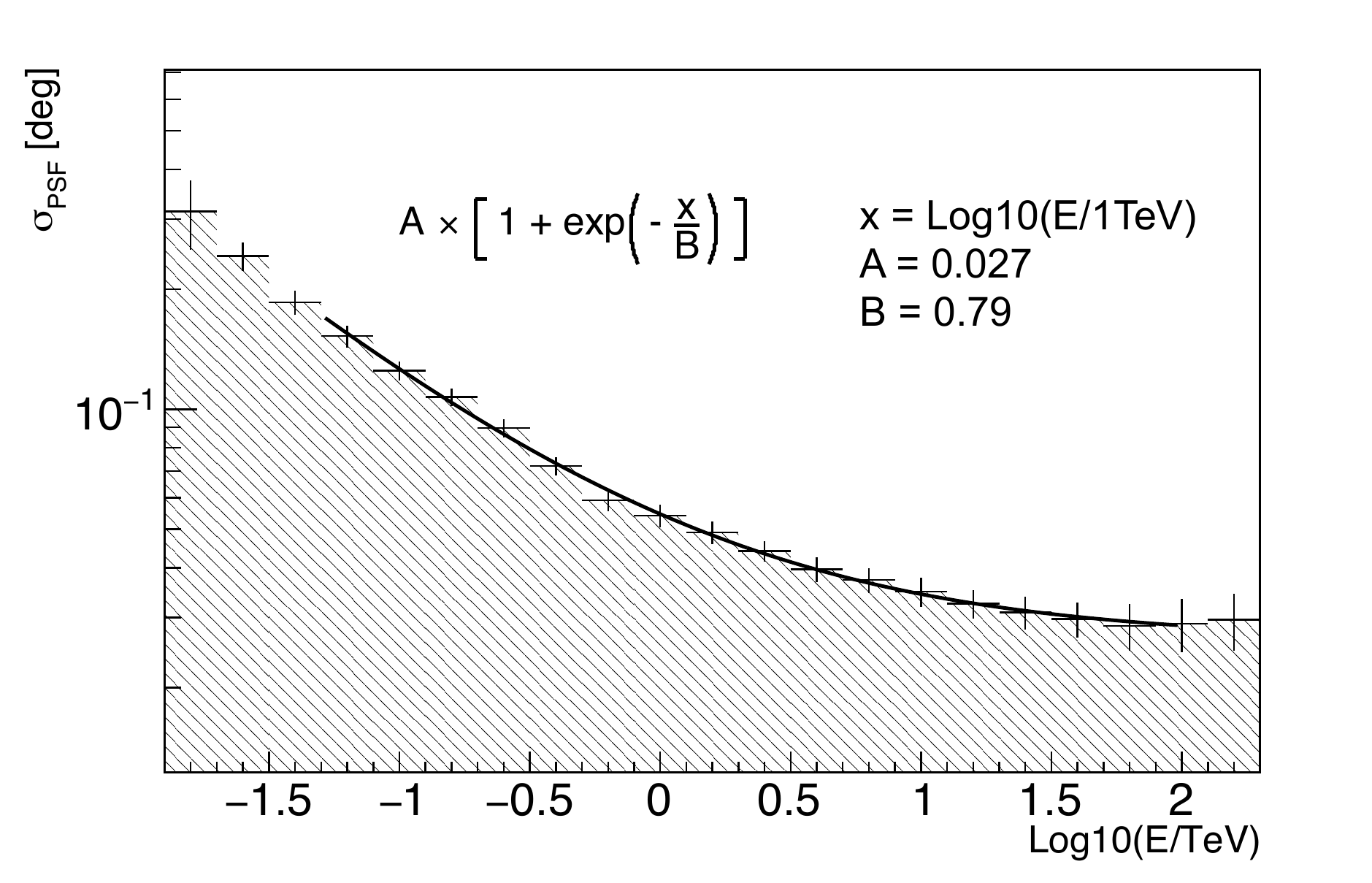}
\caption{Angular resolution of the instrument. The data are taken from the publicly available CTA performance files\footref{CTAwebpage}, solid line corresponds to the best fit. The corresponding analytical function is also shown in the plot.}
\label{fig:psf}
\end{figure}
The angular resolution as a function of the energy, $\sigma_{PSF}$, is shown in Fig. \ref{fig:psf}. Assuming $x=\log_{10}(E/\mbox{1\,TeV})$,  $\sigma_{PSF}$ can be approximated in the form: 
\begin{equation}
\label{eq:psf}
\sigma_{PSF} (x) = A \cdot \left[ 1 + \exp \left( - \frac{x}{B} \right) \right]
\end{equation}
with $A=2.71\cdot10^{-2}$\,deg representing the best angular resolution achievable with the telescope layout considered in this work and $B=7.90\cdot10^{-1}$ the scaling factor describing how fast the angular resolution changes with energy. 

The angular resolution is here defined as the angular radius that contains 68\% of the gamma-ray point spread function (PSF). For morphological studies the shape of the PSF is a key issue. To first approximation the angular resolution can be described by a two-dimensional Gaussian distribution:
\begin{equation}
\label{eq:PSF}
f_{PSF} = \exp\left( {\frac{x^2+y^2}{2\sigma_{PSF}^2}} \right)
\end{equation}
The specific values of $\sigma_{PSF}$ used in this work are provided in table \ref{tab:response}. In general, for a wide variety of telescopes operating in different energy bands of electromagnetic spectrum, in addition to the central Gaussian component, the PSF might contain tails extending well away from the peak. To account for the presence of such tails and to study the effect of their impact on the resolution of  weak gamma-ray sources, a non-Gaussian shaped PSF has been assumed. Namely, following ref. \citep{Aharonian:2006pe}, we represent the PSF in the form:
\begin{equation}
\label{eq:PSFtails}
\resizebox{.8\hsize}{!}{ $
f_{PSF} = \exp\left( {\frac{x^2+y^2}{2\sigma_{PSF}^2}} \right) + K \cdot \exp\left({\frac{x^2+y^2}{2\sigma_{PSFtails}^2}}\right) $}
\end{equation}
\begin{figure}
\centering
\includegraphics[width=8.5cm]{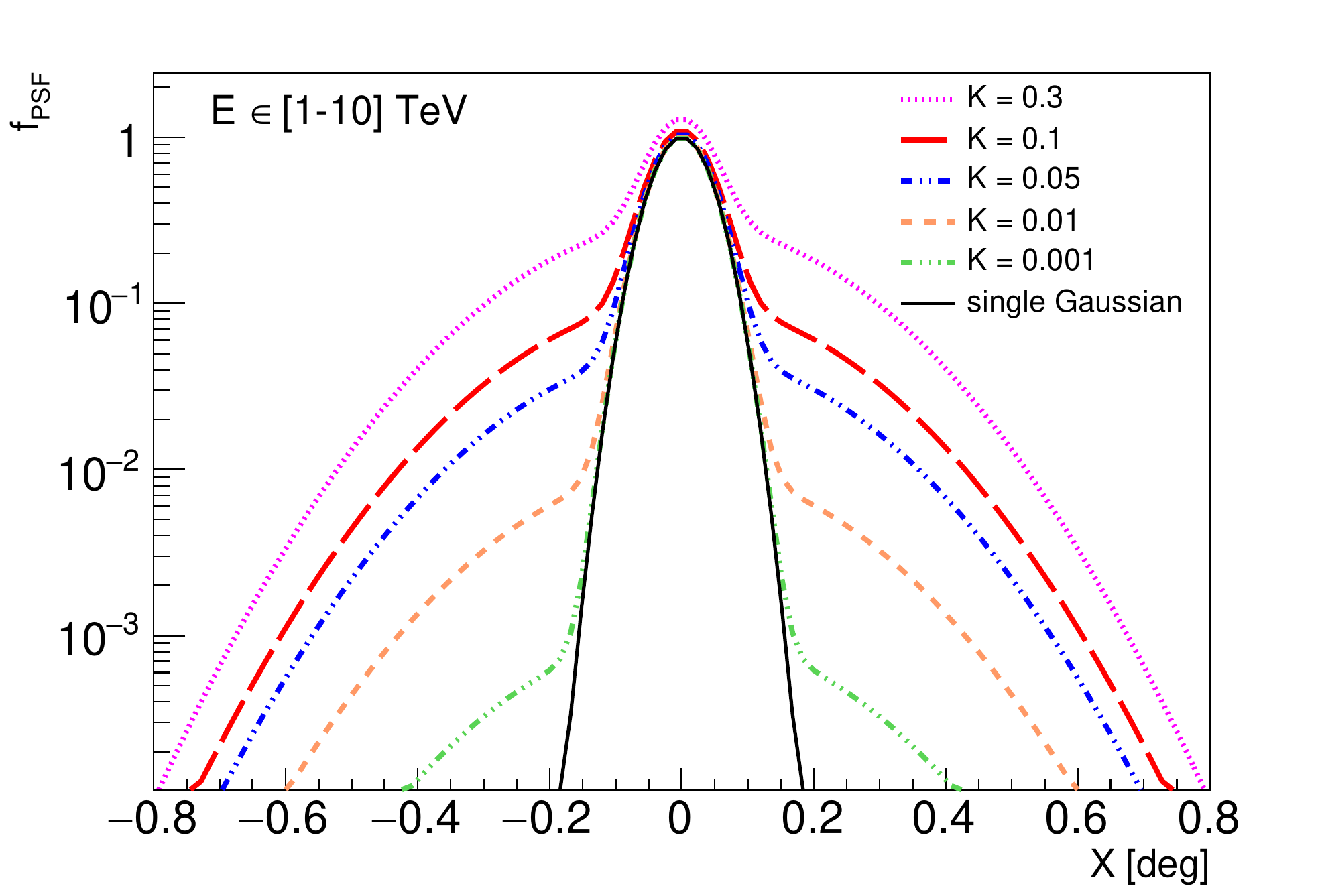}
\caption{Possible shapes of the PSF. Black curve corresponds to the ideal case of simple Gaussian PSF, described by Eq. [\ref{eq:PSF}]. Colored dashed curves are for the more realistic non-Gaussian PSF with tails, described by Eq. [\ref{eq:PSFtails}]. The curves are calculated for $\sigma_{PSF}=0.04$\,deg and five different values of the parameter $K$ (see text).}
\label{fig:PSFtails}
\end{figure} 
The tails of the non-Gaussian PSF are described by a second Gaussian function having a width $\sigma_{PSFtails}$, fixed to the fiducial value of $0.2$\,deg (assuming a worst case scenario). The ratio $K$ of the normalization factor of the main and the secondary Gaussian was adjusted to describe the effects of different tails, namely, the values considered are: 0.3, 0.1, 0.5, 0.01 and 0.001. In Fig. \ref{fig:PSFtails} the PSFs corresponding to these values are shown in the energy interval from 1 to 10\,TeV, where the instrument is expected to reach the best performance in terms of sensitivity. The black curve is for the case of single Gaussian PSF, described by Eq. [\ref{eq:PSF}]. The colored dashed curves are for the non-Gaussian PSF with tails described by Eq. [\ref{eq:PSFtails}]. Note that, even if disregarded in this study for the sake of simplicity, the location of the source within the FoV will also modify the PSF value, depending on the final configuration of the telescopes. However, this effect is negligible for energies lower than $\sim10$\,TeV, for which the PSF can be considered flat up to $4^\circ$\textbf, and only a slight degradation of the PSF (by a factor $\lesssim2$) might occur at higher energies \citep{Szanecki:2015zaa}. Moreover, it is important to highlight that the use of the HESS modeling for the non-Gaussian PSF, as reported in ref. \citep{Aharonian:2006pe}, has to be considered here as a conservative upper limit. With tens of telescopes, the CTA observatory is expected to do better and this especially concerns the observations at the high energies, i.e. $E\ge10$\,TeV. In fact, although a constant value of $\sigma_{PSFtails}$ is assumed in this work by virtue of the HESS results, a reduction of the tails size is likely to take place as the energy increases, due to the larger telescope multiplicity which should reduce the PSF fluctuations responsible for the tails. The energy dependence of the PSF tails is beyond the goals of this paper. Nevertheless, detailed MC studies aimed to explore the effect of the tails on the instrument performance and their relevance in different energy domains might represent an extremely important task for the upcoming CTA observatory.

\subsection{Effective Area} 
\label{cap:Aeff}

\begin{figure}
\centering
\includegraphics[width=8.5cm,height=5.8cm]{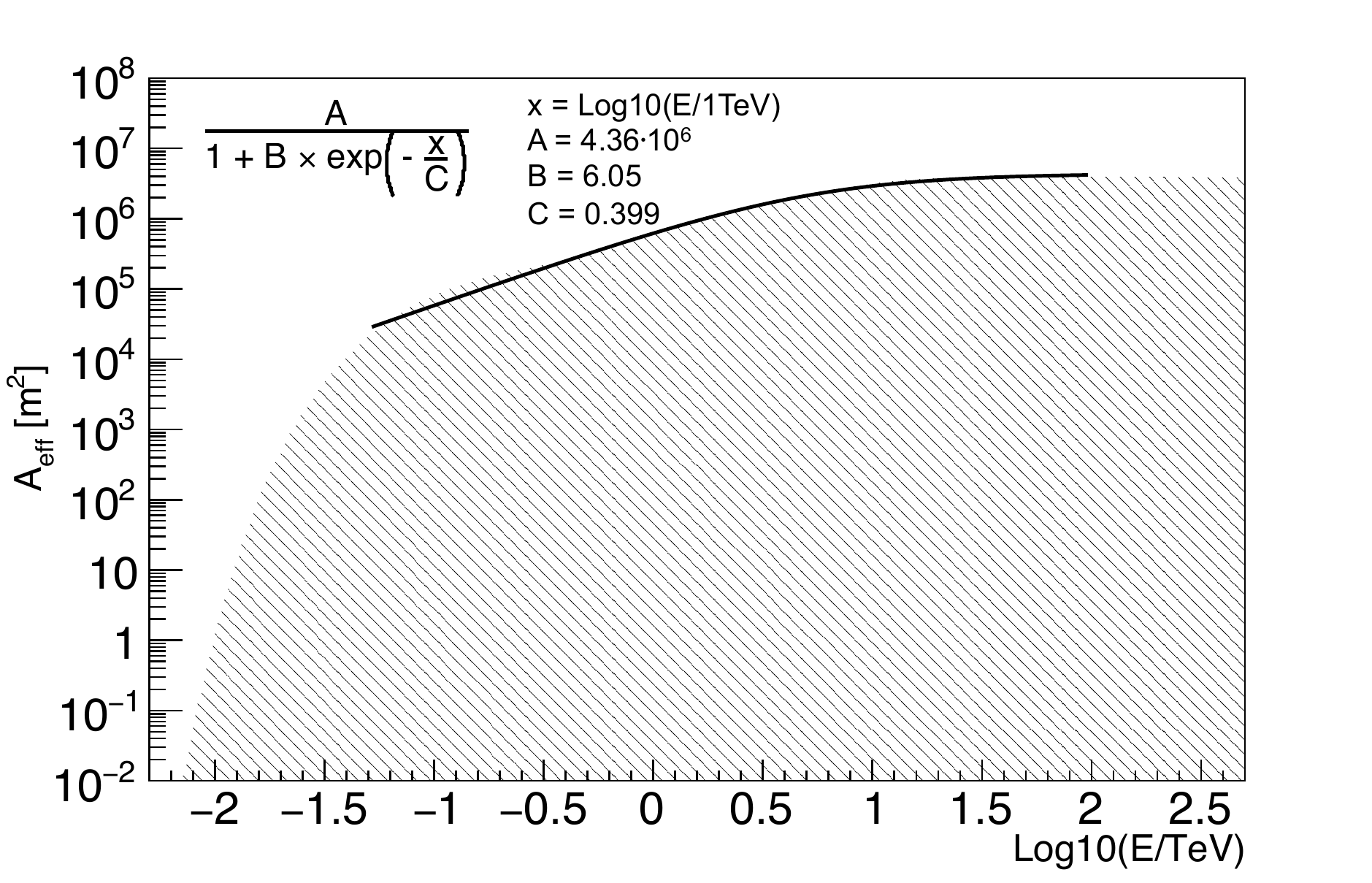}
\caption{Effective area of the telescope layout considered in this study. The data are taken from publicly available calculations of the CTA performance\footref{CTAwebpage}, solid lines correspond to the parametrization formula which is also shown in the figure.}
\label{fig:aeff}
\end{figure}
While for a single telescope the effective area is determined by the radius of the Cherenkov light pool at ground ($r_{light-pool}\sim120$\,m), the effective detection area of a multi-telescope system is determined essentially by the total geometrical area \citep{Aharonian1997343}, as can be seen in Fig. \ref{fig:aeff}. For the considered layout of the CTA South observatory, the effective detection area $A_{eff}$ can be parametrized with the following expression in the energy range from 50\,GeV to 100\,TeV:
\begin{equation}
\label{eq:aeff}
A_{eff} (x) = \frac{A}{1 + B \cdot \exp \left( -\frac{x} {C} \right) } 
\end{equation}
where the saturation value of the effective area is $A=4.36 \cdot 10^6$\,$\mbox{  m}^2$, while $B=6.05$ and $C=3.99\cdot10^{-1}$ define the rate of change of $A_{eff}$ with respect to energy. 

\subsection{Energy resolution} 
\label{cap:Eres}

\begin{figure}
\centering
\includegraphics[width=8.5cm,height=5.8cm]{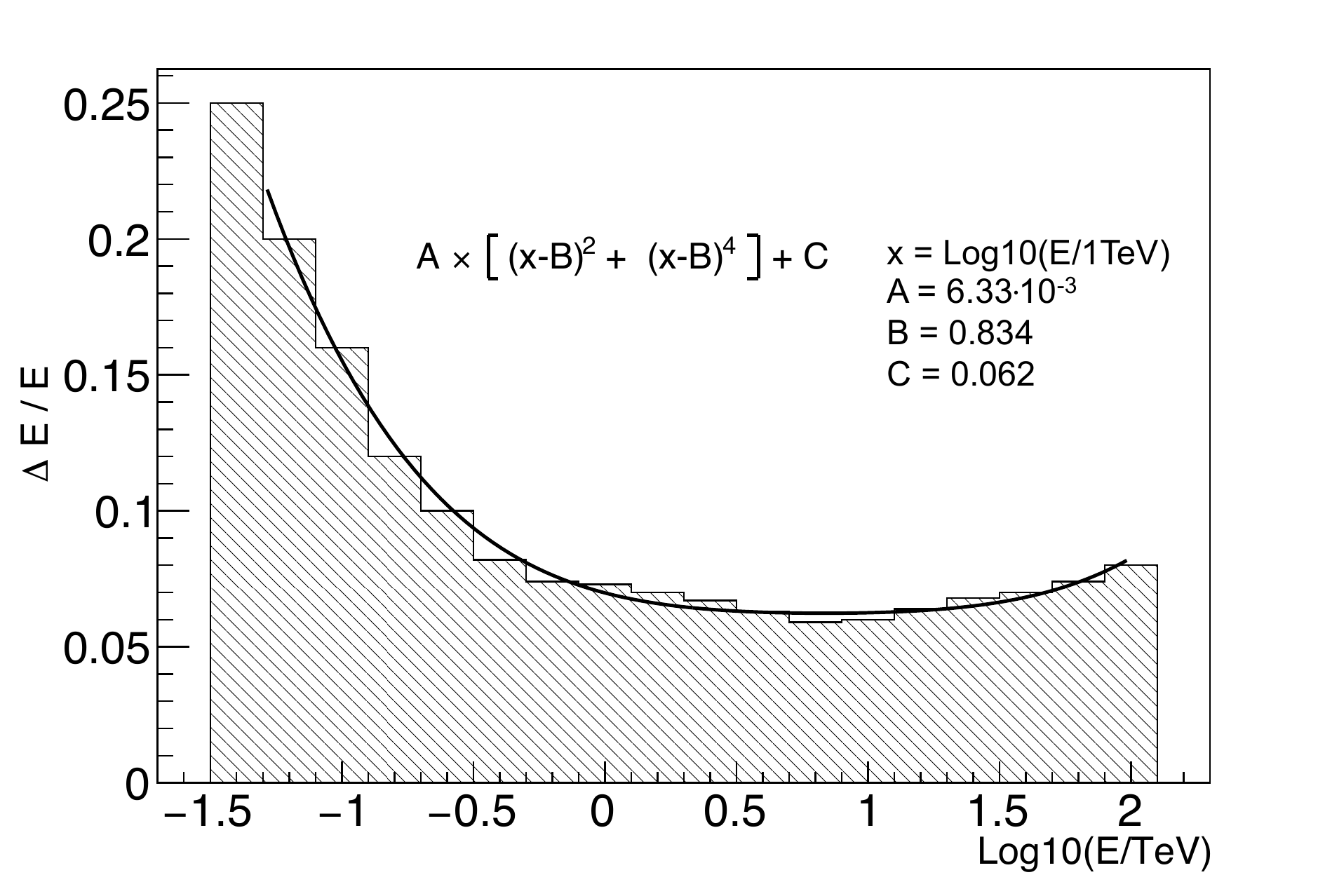}
\caption{Energy resolution for the considered layout of CTA southern array. The data are taken from publicly available calculations of the CTA performance\footref{CTAwebpage}. The solid line corresponds to the analytical parametrization which is also shown in the figure.}
\label{fig:Eres}
\end{figure}

The energy resolution of the instrument, referring to the public CTA MC results, is shown in Fig. \ref{fig:Eres}. Although energy resolution is rather modest in the low energy range (approximately 20\% around 50 GeV), it significantly improves at higher energies saturating at the level of 6 to 8\% from 1\,TeV to 100\,TeV. The energy dependence of the resolution can be parametrized in the following form:
\begin{equation}
\label{eq:eres}
\left( \Delta E / E  \right) (x) =  A \times \left[ (x-B)^2 + (x-B)^4 \right] + C
\end{equation}
with the normalization factor $A$ taking the value $A=6.33\cdot10^{-3}$, the parameter $B=8.34\cdot10^{-1}$ fixing the value of the energy for which the resolution takes its best value and $C=6.24\cdot10^{-2}$ representing the best energy resolution achievable with the telescope layout considered in this work. A detailed study of the energy resolution effects on the observation potential of a CTA-like instrument is beyond the framework of this paper and will be addressed in a future work. 

\subsection{Background rate} 
\label{cap:bkgrate}

\begin{figure}
\centering
\includegraphics[width=8.5cm,height=5.8cm]{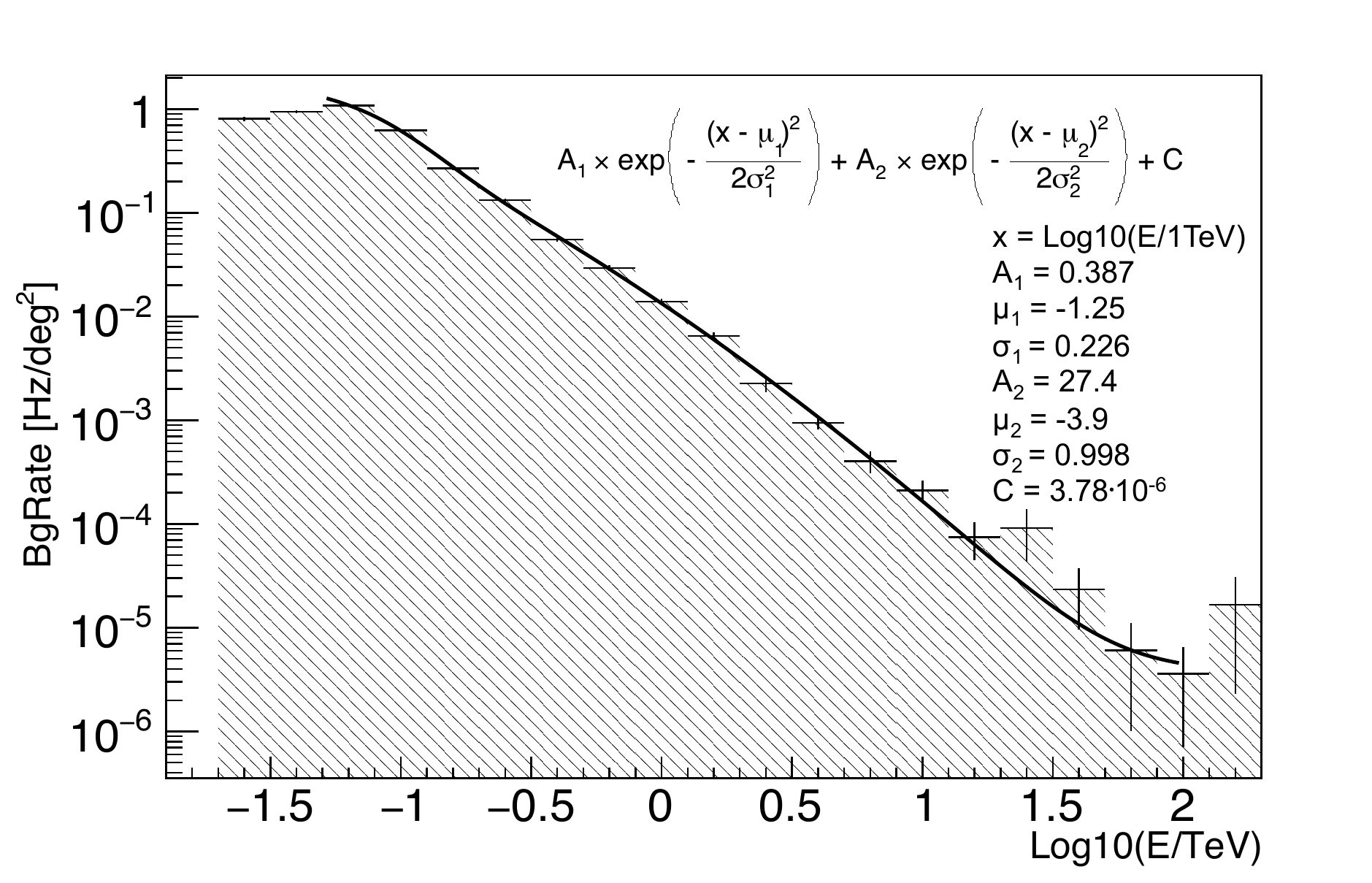}
\caption{Background rate per unit of solid angle. The data are taken from publicly available calculations of the CTA performance\footref{CTAwebpage}. The solid line corresponds to the parametrization formula which is also shown in the figure.}
\label{fig:bgrate}
\end{figure}

The simulated rate of background events after all the rejection cuts that we used in our study is shown in Fig. \ref{fig:bgrate} as function of the energy. Simulations were set up by the CTA Consortium assuming power-law spectra, except for the electron and positron background, the latter being described by a log-normal peak on top of an $E^{-3.21}$ power-law spectrum. For the background rate initiated by protons and nuclei of cosmic rays, an $E^{-\alpha}$ dependence has been adopted, with $\alpha$ ranging from $2.70$ for protons to $2.63$ for iron, see e.g. ref. \citep{2013APh....43..171B}. The noise from the night-sky background ($\sim100$\,MHz for $100$\,m$^2$ dish and $0.15^\circ$ pixel\footnote{The value for the night-sky background can be checked at: \url{https://www.stfc.ac.uk/files/photosensors-and-electronics-for-cta/}.}, corresponding to dark-sky observations towards an extra-Galactic field  \citep{Maier:2015bta}) and from electronics is added as well. On the simulated background showers, selection cuts are then applied in order to suppress events not induced by gamma-rays (for the details on the analysis performed by the CTA Collaboration see e.g. \cite{Hassan:2015bwa}). For the surviving events, the energy dependence of the overall background rate can be approximated to the following form:
\begin{equation}
\label{eq:bgrate}
\resizebox{.85\hsize}{!}{ $
BgRate (x) = A_1 \cdot \exp \left( - \frac{(x-\mu_1)^2}{2\cdot\sigma_1^2}\right) + A_2 \cdot \exp \left( - \frac{(x-\mu_2)^2}{2\sigma_2^2}\right) + C 
$ }
\end{equation}
with $A_1=3.87\cdot10^{-1}$\,Hz/deg$^2$, $\mu_1=-1.25$, $\sigma_1=2.26\cdot10^{-1}$, $A_2=27.4$\,Hz/deg$^2$, $\mu_2=-3.90$, $\sigma_2=9.98\cdot10^{-1}$ and $C=3.78\cdot10^{-6}$\,Hz/deg$^2$.

For each energy interval, we computed the mean number of spurious events $N_{BgRate}$ scaling the rate $BgRate$ by the observation time and by the angular area. In order to take into account fluctuations in the background, we randomly sampled the number of background events $N_{B}$ from a Poissonian distribution: $f(N_{B} \mid \lambda) = (\lambda^{N_B} e^{-\lambda})/{N_B!}$, with expected value $\lambda = N_{BgRate}$.

\subsection{Sensitivity} 
\label{cap:CTAsensitivity}

The sensitivity of the telescope is defined as the minimum flux of gamma rays required for a statistically significant detection. The publicly available CTA sensitivity curve is obtained from detailed MC simulations with the baseline analysis being applied to simulated data \citep{Hassan:2015bwa}, calculated for the observation of a point-like source with a Crab-like energy spectrum. For the calculation of the instrument sensitivity one needs to specify the requirements to accept a signal as statistically significant in a certain observation time scale. In the case of the ground-based gamma-ray instruments it is usually reduced to two conditions (see e.g.  \citep{Aharonian1997343}): (i) the presence of at least 10 excess events and (ii) a significance of at least five standard deviations calculated according to the formulation of \emph{Li \& Ma} \citep{lima}.  In addition, in order to account for the background systematics, a minimum signal excess over the background uncertainty is required. Regarding this last point, the CTA prescription is to assume an accuracy of the background modeling and subtraction of 1\% of the remaining background events, and to require a signal excess at five times this 1\% background systematic uncertainty.
In summary, for the computation of the differential sensitivity, the following conditions have been applied by the CTA Consortium for each energy interval (five intervals per decade of reconstructed energy, i.e. intervals of the decimal exponent of 0.2):
\begin{itemize}
\item $S\ge 10$
\item $N_{\sigma}\ge 5$
\item $S/B \ge 0.05$
\end{itemize}
where $S$ and $B$ denote the signal and background events in the source region, which in the \emph{Li \& Ma} notation correspond to $N_{on} = S + B$ and $N_{off} = B/\alpha$, with $\alpha$ being the ratio of the on-source time to the off-source time during which $N_{on}$ and $N_{off}$ photons are counted, respectively, and which also depends on the different integration regions used to estimate the two.

\section{Morphological studies} 
\label{singlesource}

In this section we start from simulations of an isolated source and use them as a benchmark to test the instrument performance for different observation modes. First the event rates and the background regimes are investigated. Then the morphological reconstruction of the isolated source, aimed to estimate its center of gravity and its angular size, is evaluated assuming both Gaussian and non-Gaussian PSFs. To reconstruct the signal ($S$), a $\chi^2$-fitting is conservatively used here for simplicity; more sophisticated approaches (see e.g. \citep{Knodlseder:2014bka, Becherini:2012iy}) might improve the results shown below.

\subsection{Isolated source simulation} 
\label{}

For simulating the source we created an excess map of $3^\circ \times 3^\circ$ with pixel size of $0.03^\circ$. The map was filled with both signal and background events. The background events were uniformly distributed in the map, whereas the gamma-ray source was simulated assuming a Gaussian shape described as:
\begin{equation}
\label{eq:gaussianshape}
\resizebox{.85\hsize}{!}{ $f(x,y) = A\cdot exp\left( - \left( \frac{(x-X_0)^2}{2\sigma_{src}^2} \right) + \left( \frac{(y-Y_0)^2}{2\sigma_{src}^2} \right) \right)$ }
\end{equation}
centered on the point $(X_0,Y_0)=(0,0)$\,deg and characterized by the size $\sigma_{src}$. The normalization factor $A$ takes into account the strength of the gamma-ray source. Concerning the size of the source, three different scenarios have been investigated:
\begin{itemize}
	\item $\sigma_{src}=0.03$\,deg; this is smaller than the PSF in any energy band, therefore the source can be considered as a point-like source;
	\item $\sigma_{src}=0.1$\,deg; this is comparable to the PSF,  therefore the source can be considered as a moderately extended source;
	\item $\sigma_{src}=0.2$\,deg; this is larger than the PSF ($\sim 2$ times),  therefore the source can be considered as an extended source.
\end{itemize}
In each energy interval and for each of these three angular scales, we simulated gamma-ray detection assuming the following form for the gamma-ray flux:  
\begin{equation}
\label{eq:CrabFlux}
 \frac{dN}{dE} =  n\cdot N_0\times \left( \frac{E}{1\,TeV} \right)^{-\alpha}
\end{equation}
with $\alpha=2.62$ which corresponds to the Crab power-law spectrum as measured by HEGRA \citep{Aharonian:2004gb}. The flux strength is given in units of Crab flux at 1\,TeV: $N_0=2.83\cdot 10^{-11}\,\mbox{TeV}^{-1}\mbox{cm}^{-2}\mbox{s}^{-1}$. 

\subsection{Event rates and background regimes} 
\label{detectionrate}
\begin{figure*}[!htp]
     \begin{center}
     \subfigure{%
     \includegraphics[width=\columnwidth]{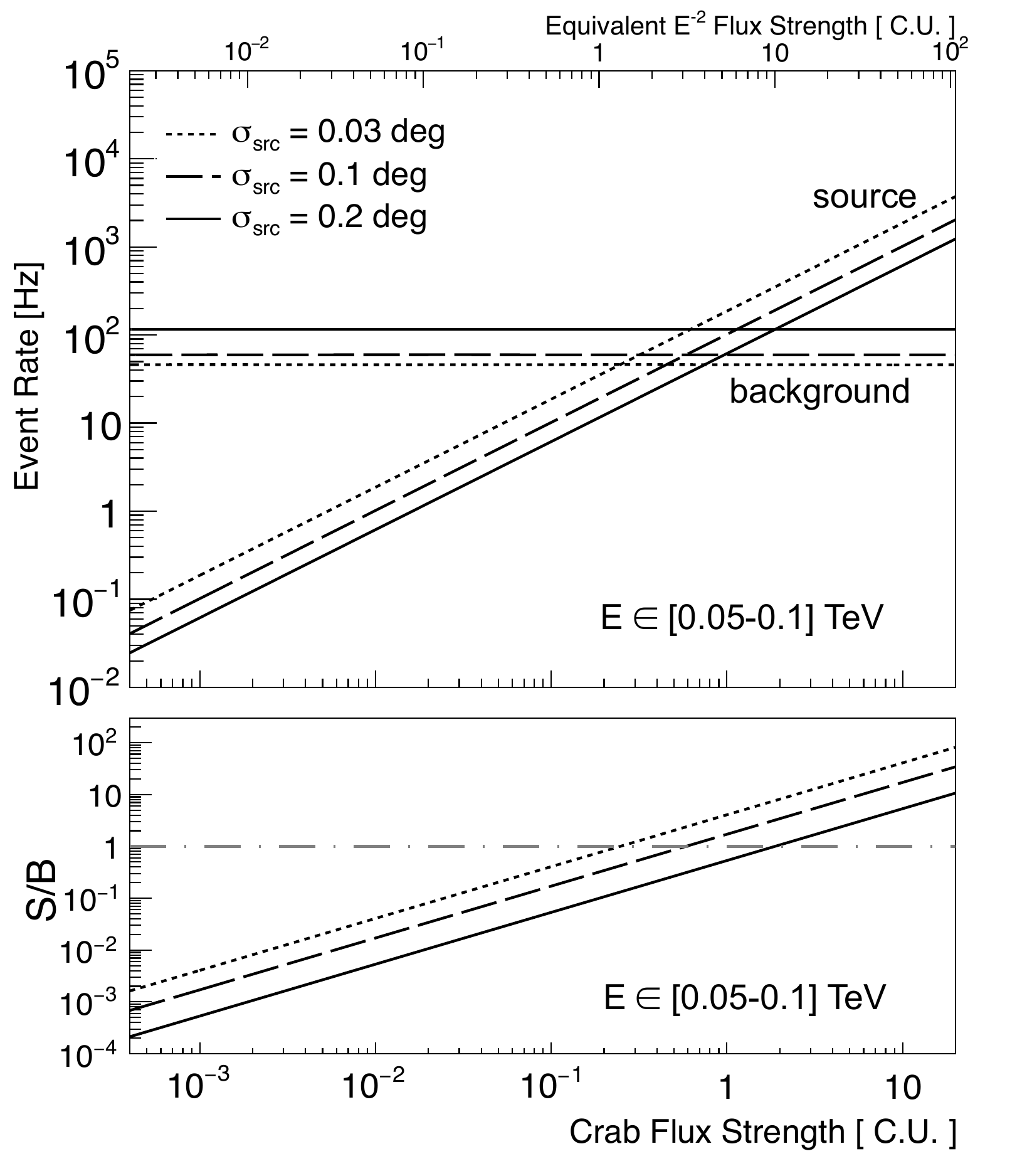}
      }
      \subfigure{%
      \includegraphics[width=\columnwidth]{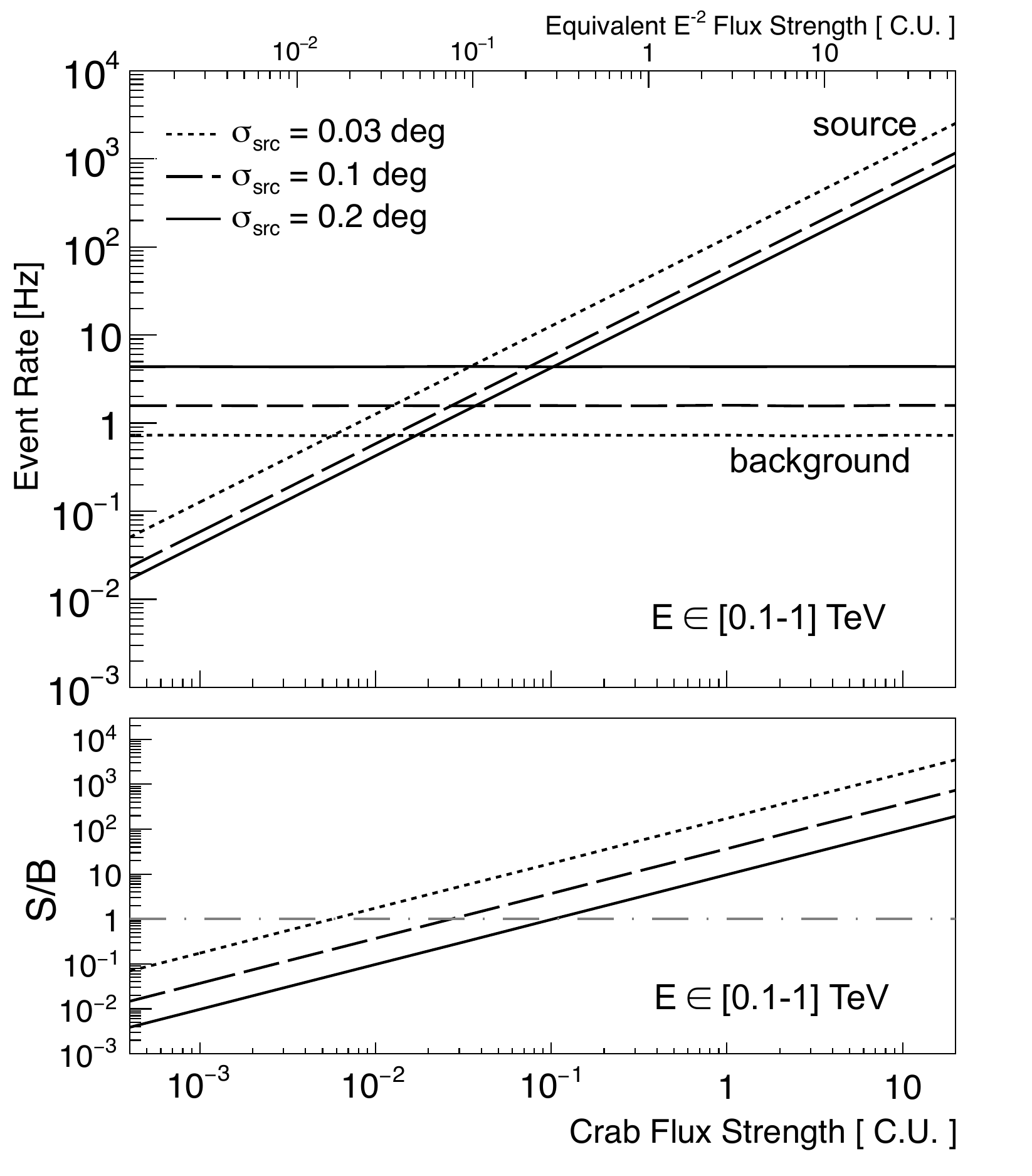}
      } \\
      \subfigure{%
      \includegraphics[width=\columnwidth]{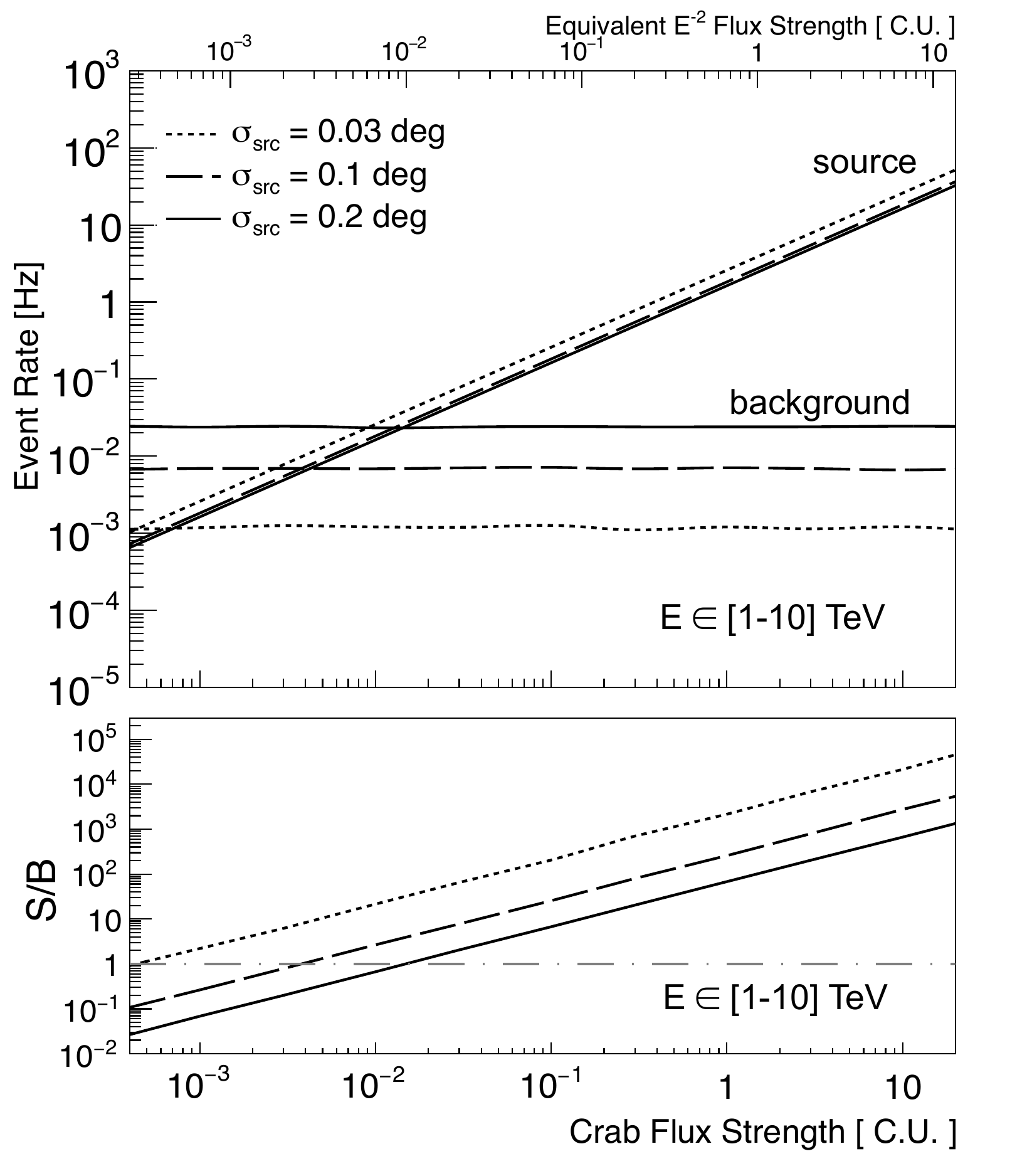}
      }
      \subfigure{%
      \includegraphics[width=\columnwidth]{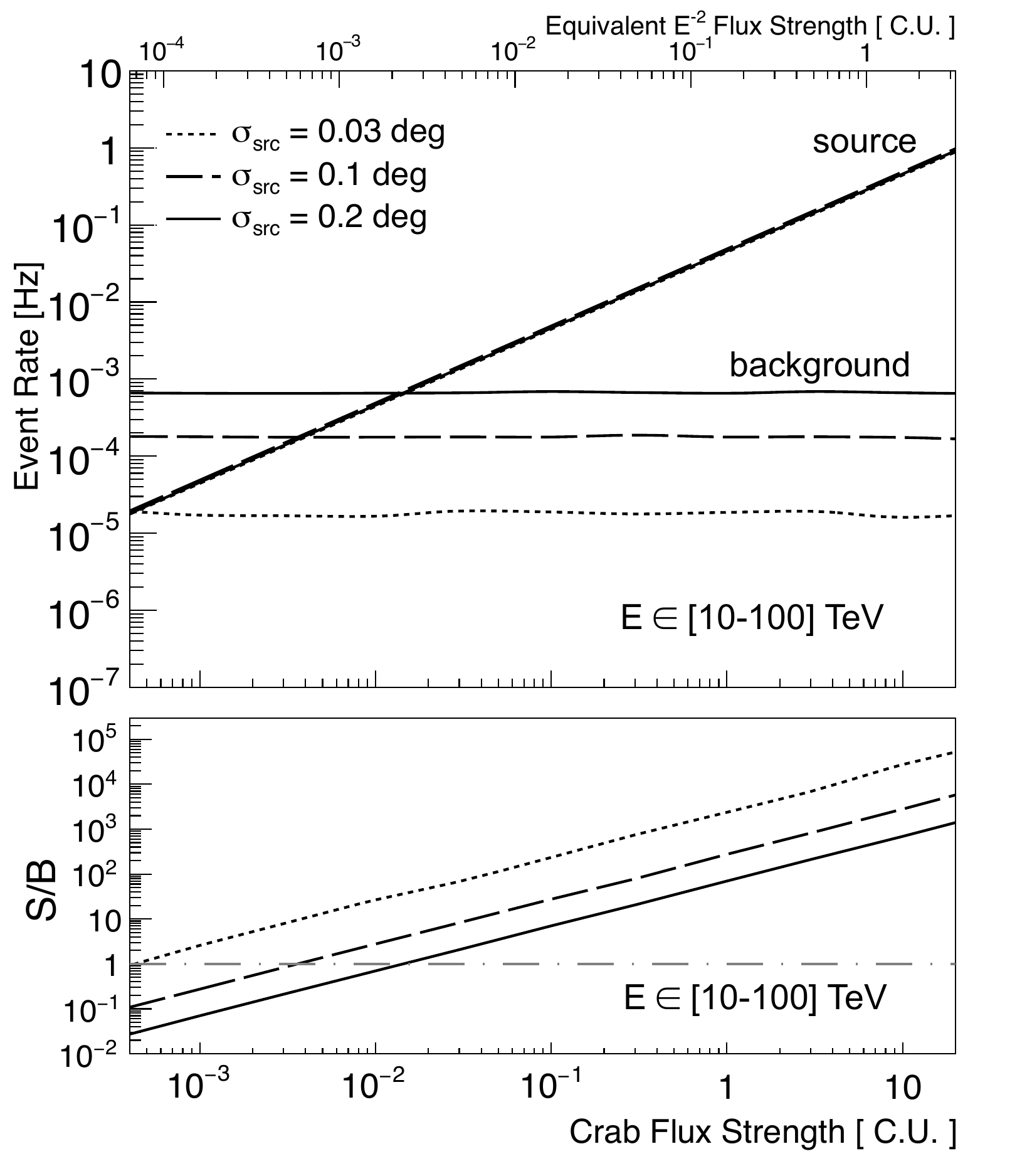}
      }
    \end{center}
\caption{Event rates for background events ({\it background, B}) and signal events ({\it source, S}). The ratio of the two ($S/B$) is also shown at the bottom of each panel; the gray dashed-dotted line is for $S/B=1$. The four panels are for the four different energy ranges, which are shown in each figure. The three sets of curves correspond to three sizes of the source: $0.03$\,deg (dotted lines), $0.1$\,deg (dashed lines) and $0.2$\,deg (solid lines). The event rates are calculated for different ROI of the signal, depending on the source size (see text for more details).}
\label{fig:SB}
\end{figure*}

In Fig. \ref{fig:SB} the rates for the signal $S$ and for the background $B$ events as well as the ratio $S/B$ are shown for different source flux strengths and source sizes. To estimate the $S$ and $B$ in the {\itshape region of interest} (ROI), we derive the number of photons from a region on the signal and background maps respectively, defined as a circle centered on the center of the source and having radius $r_{ROI} = \sqrt{ (2\sigma_{src})^2 + (2\sigma_{PSF})^2}$, which in the case of point-like objects is reduced to $R_{ROI} = 2\sigma_{PSF}$. On the upper horizontal axis of each plot the corresponding flux levels for an $E^{-2}$ power-law spectrum are shown in units of Crab at 1\,TeV. Although the source regions have been filled with the same number of events for each assumed flux strength, and the flux is uniformly spread all over the source extension, in the low energy intervals the signal rate $S$ slightly differs depending on the source size. This is due to the larger $\sigma_{PSF}$ at these energies, which spreads out events on a larger scale, pushing events that belong to the tails of the source distribution to fall outside the ROI.

When $S/B<1$, the detection proceeds in the {\itshape background dominated} regime. For larger values (presented in table \ref{tab:SB} for each energy interval) the detection proceeds in the {\itshape background free} regime. The integrated flux in the region comprised by sources of different size is normalized to the same value (or percent of the Crab flux). Therefore, to achieve the $S/B=1$ condition, the required flux is higher for larger source sizes.
\begin{table}[htbp]
\begin{center}
\resizebox{\linewidth}{!}{
\begin{tabular}{ | c | c | c | c |}
  \hline 
	Energy				&	$\sigma_{src}=0.03$\,deg		&		$\sigma_{src}=0.1$\,deg		&		$\sigma_{src}=0.2$\,deg   \\
  \hline
  $[0.05-0.1]$\,TeV	& 	$0.3$ C.U.	 	&  		$0.5$ C.U. 		&  		$1$ C.U. 	\\
  $[0.1-1]$\,TeV 	& 	$0.006$ C.U.		& 		$0.03$ C.U.  		& 		$0.1$ C.U.  \\
  $[1-10]$\,TeV  	& 	$<0.001$ C.U. 		& 		$0.004$ C.U.  		& 		$0.01$ C.U.  \\
  $[10-100]$\,TeV 	& 	$<0.001$ C.U. 		& 		$0.004$ C.U.  		& 		$0.01$ C.U.  \\
  \hline  
\end{tabular}
}
\caption{\label{tab:SB}The flux strengths in Crab Units (C.U.) for which the condition $S/B=1$ is reached for different source sizes in four energy intervals.}
\end{center}
\end{table}

\subsection{Signal-to-noise} 
\label{signaltonoise}

\begin{figure}[htbp]
\centering
\includegraphics[width=8.2cm]{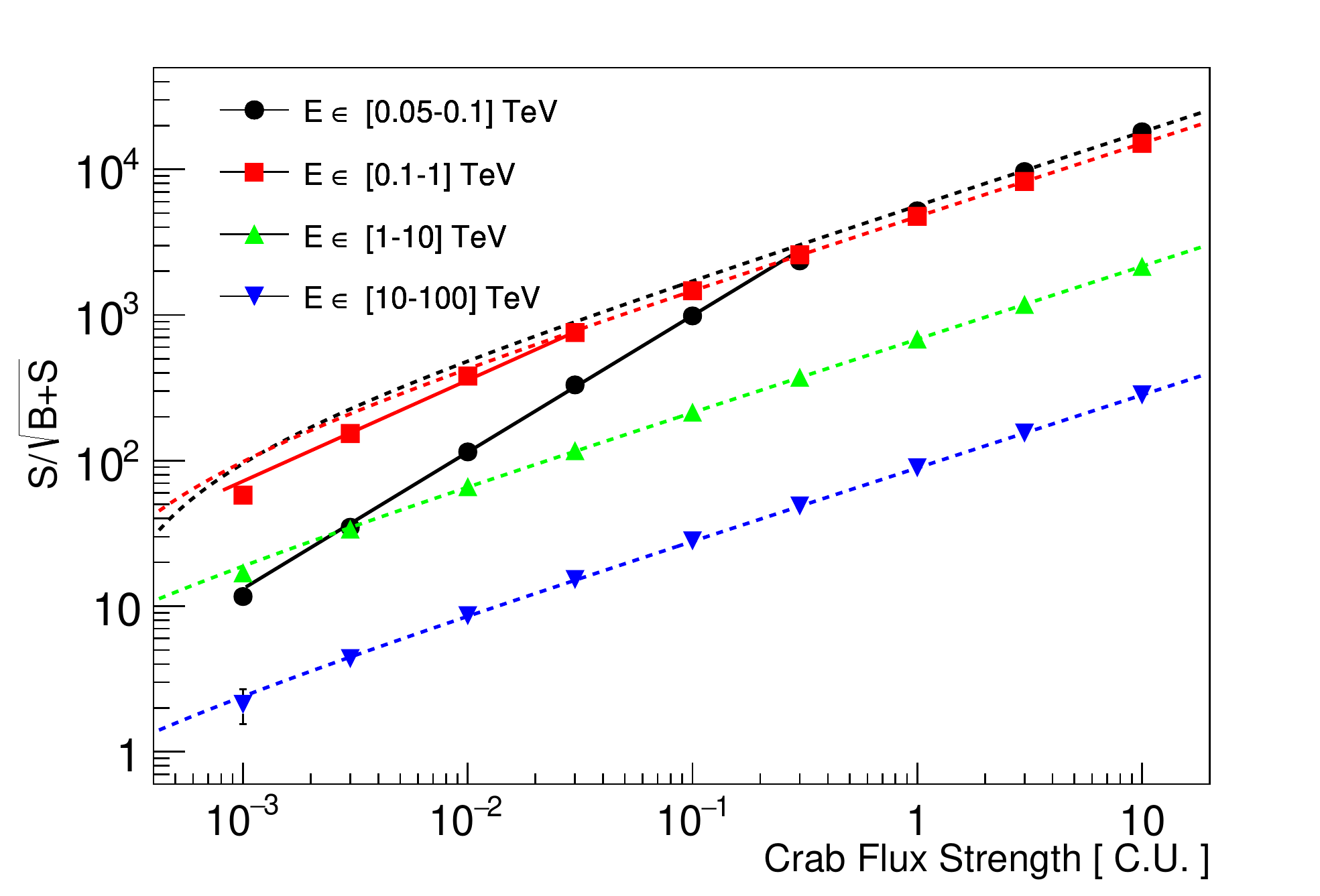}
\caption{Signal-to-noise ratio as a function of C.U. flux (for a point-like source emitting photons according to the Crab-like energy spectrum and observed for 50\,h). Four different energy domains are shown: (0.05-0.1)\,TeV in black, (0.1-1)\,TeV in red, (1-10)\,TeV in green and (10-100)\,TeV in blue.}
\label{fig:signalnoise}
\end{figure}

In addition to the requirement of adequate statistics of the signal events, a sufficient excess over the background level, i.e. {\itshape signal-to-noise ratio} ($S/N$, where $N=\sqrt(S+B)$), is requested to ascertain the reliability of detection. 
The signal-to-noise ratio can be described in terms of the event rates:
\begin{equation}
\frac{S}{N}=\frac{T\cdot R_S}{\sqrt{T\cdot R_S + T\cdot R_B}}
\end{equation}
where $T$ is the observation time and $R_S$ and $R_B$ are the event rates for signal and background, respectively. For a given observation time, when $R_B \gg R_S$, then $S/N \propto R_S$ and  when $R_B \ll R_S$, then $S/N \propto \sqrt{R_S}$. Thus, we expect a linear dependence of $S/N$ as a function of the flux strength in the background dominated regime, and a square root dependence when the signal dominates over the background. This trend can be seen in Fig. \ref{fig:signalnoise}.

\subsection{Reconstruction of morphological parameters for Gaussian PSF response}  
\label{cap:recosignlesource}

A Gaussian-shaped source convolved with the instrument PSF (described also with a Gaussian function) is used to fit the reconstructed excess map:
\begin{equation}
\label{eq:recofunc}
\resizebox{.8\hsize}{!}{ $f(x,y) = A\cdot exp\left( - \left( \frac{(x-X_0)^2}{2(\sigma_{src}^2+\sigma_{PSF}^2)} \right) + \left( \frac{(y-Y_0)^2}{2(\sigma_{src}^2+\sigma_{PSF}^2)} \right) \right)$ }
\end{equation}
where $\sigma_{PSF}$ depends on the energy interval (see table \ref{tab:response}). Using a sample of tens of realizations of the simulation and treating $\sigma_{PSF}$ as a fixed parameter, we can estimate the size $\sigma_{src}$ and the center of gravity ($X_0$, $Y_0$) by performing a $\chi^2$-fitting analysis to the Gaussian shaped spatial distribution of the source. 

\begin{figure}[htbp]
     \begin{center}
     \subfigure{%
     \includegraphics[width=.93\columnwidth]{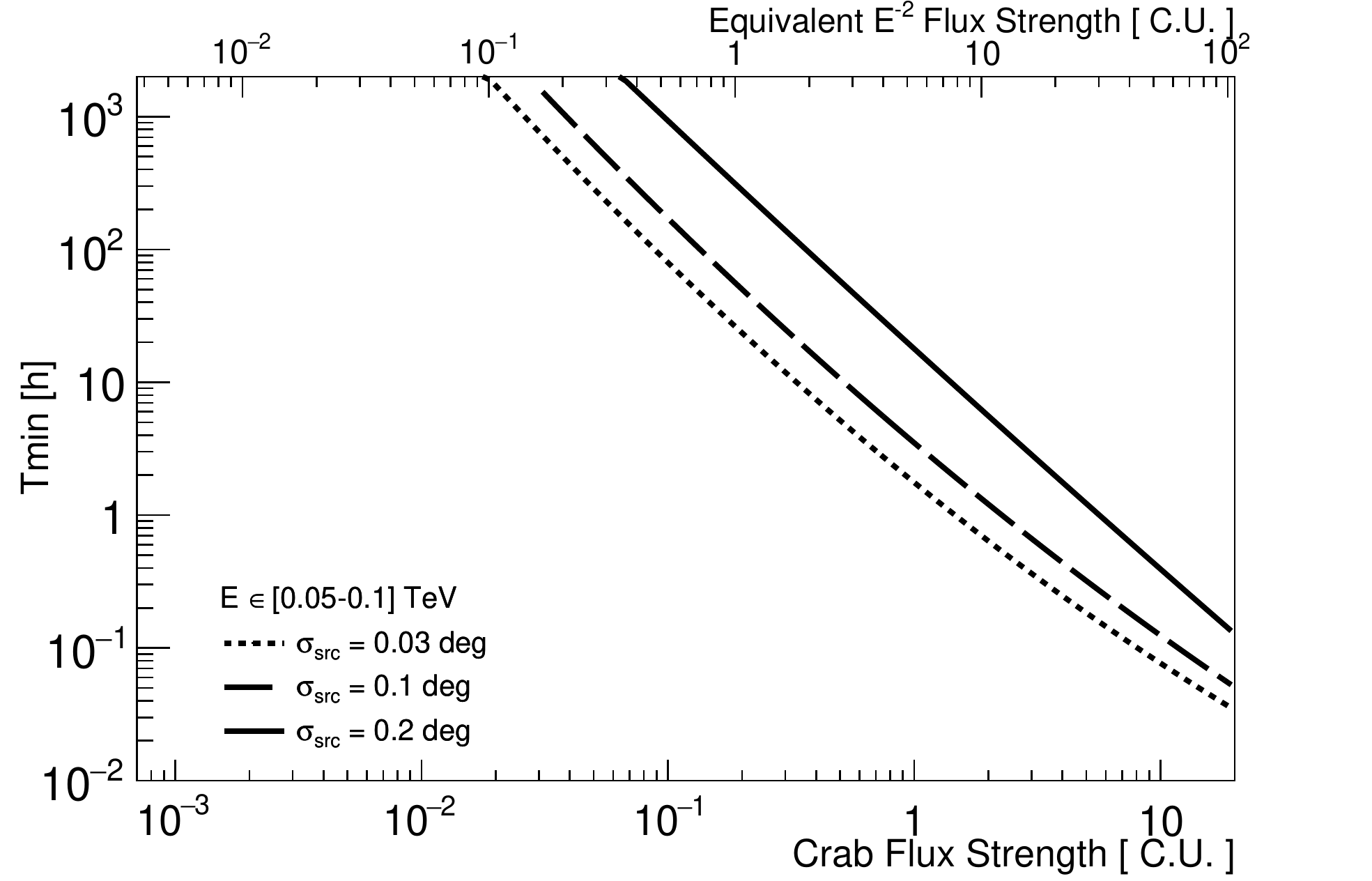}
      } \\
      \subfigure{%
      \includegraphics[width=.93\columnwidth]{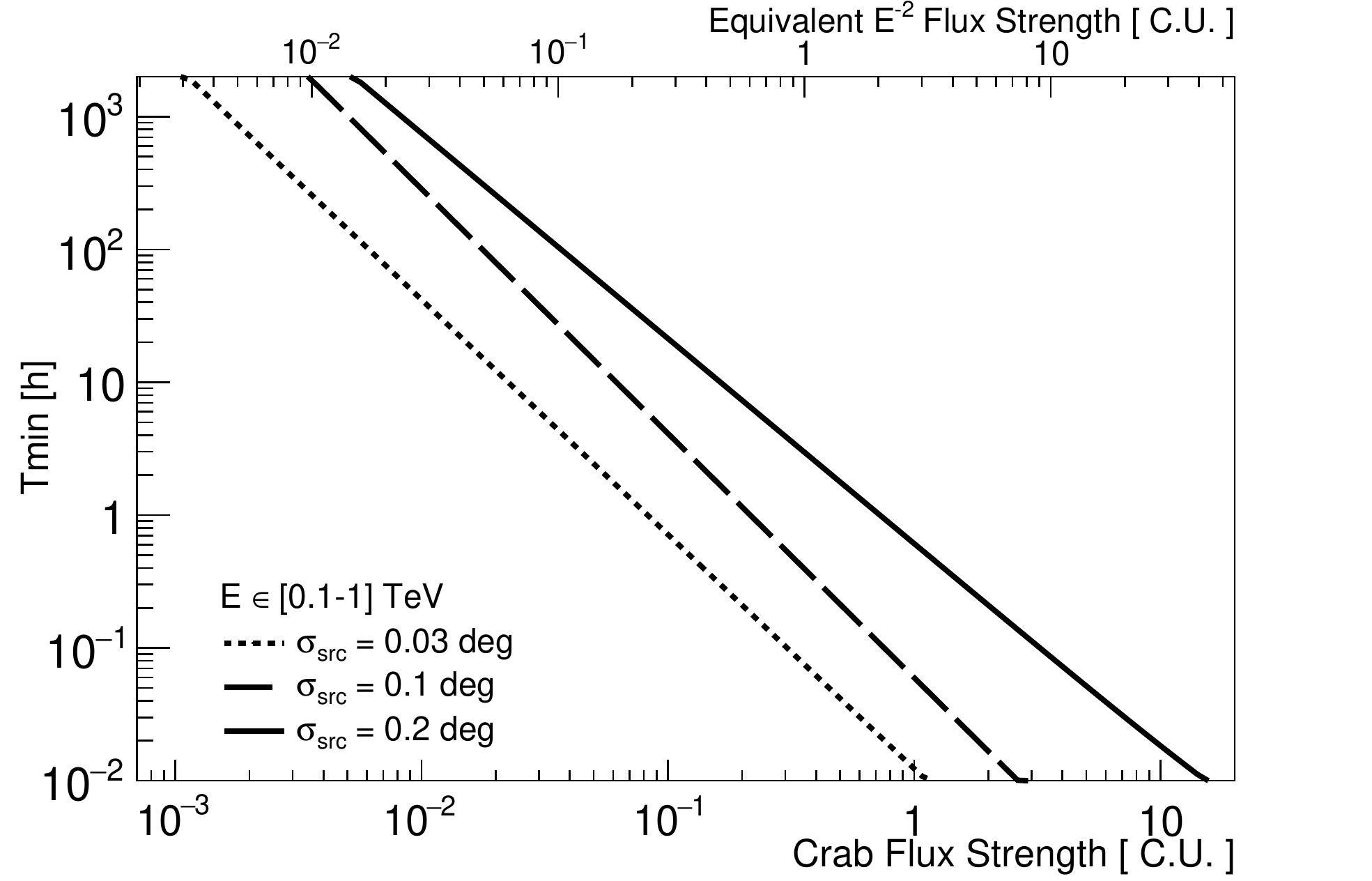}
      } \\
      \subfigure{%
      \includegraphics[width=.93\columnwidth]{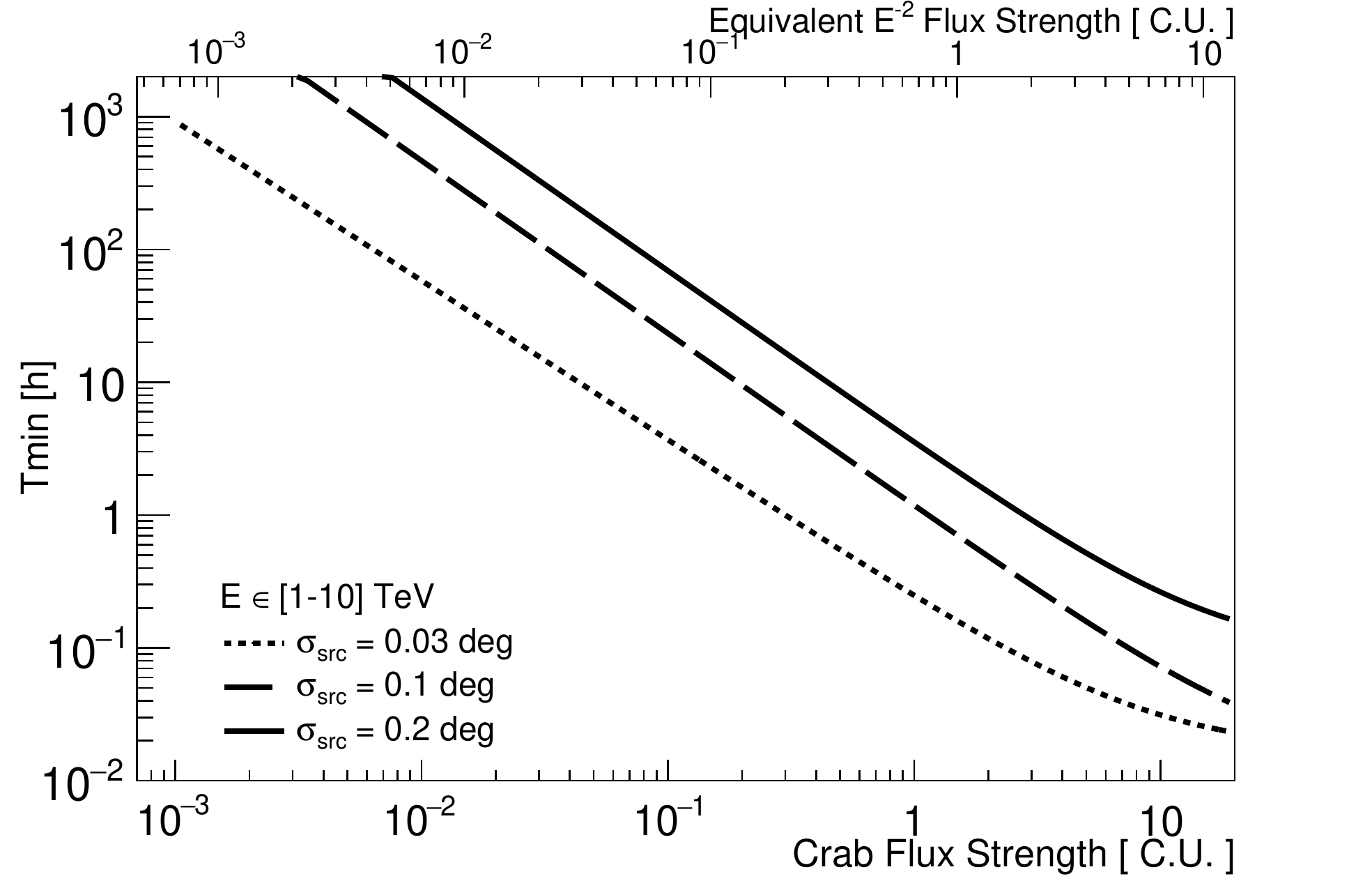}
      } \\
      \subfigure{%
      \includegraphics[width=.93\columnwidth]{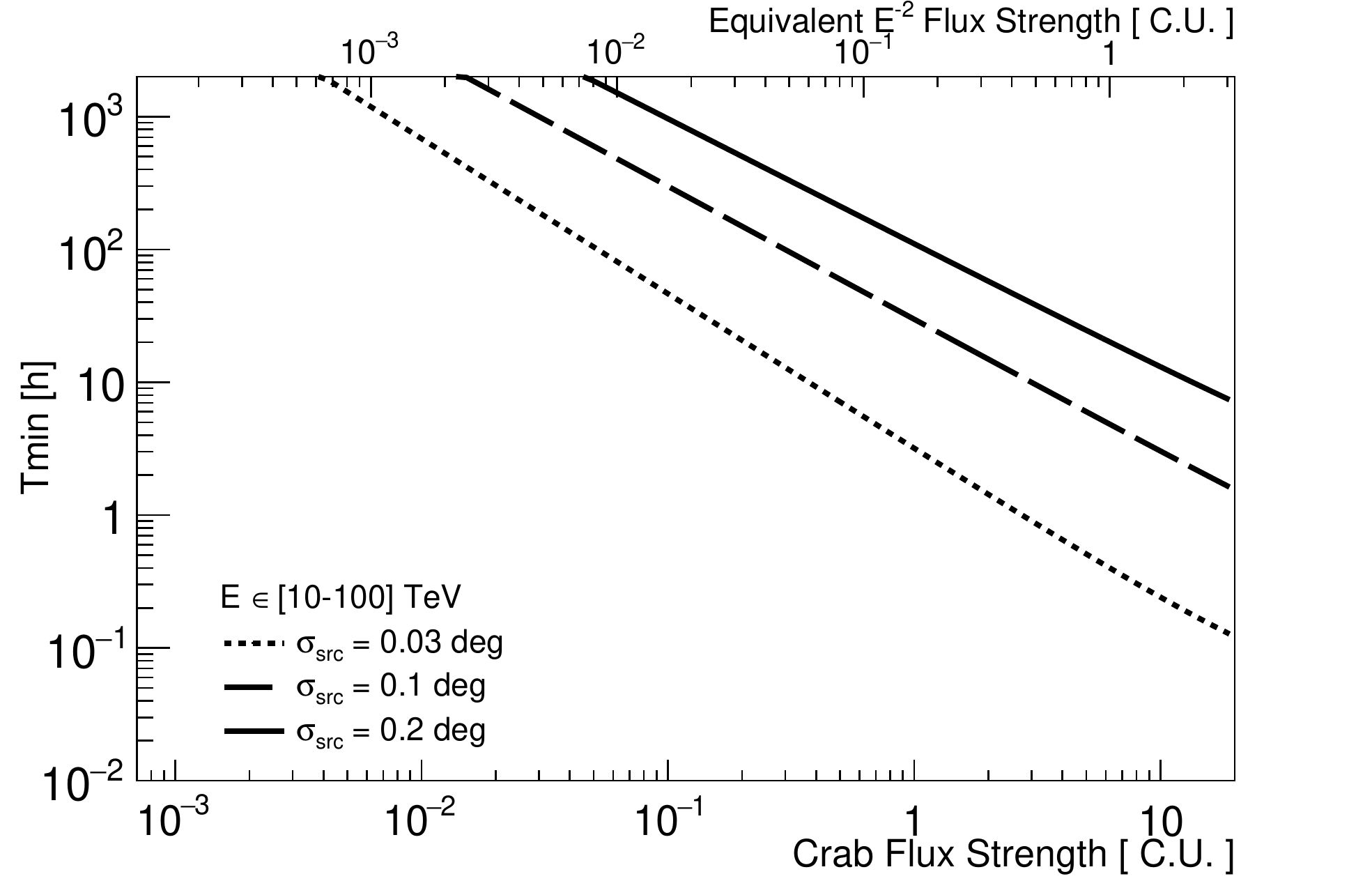}
      }
    \end{center}
\caption{Minimum time needed to reconstruct the center of gravity of the source in the case of Gaussian PSF response and for four energy intervals: $[0.05-0.1]$\,TeV, $[0.1-1]$\,TeV, $[1-10]$\,TeV and $[10-100]$\,TeV.}
\label{fig:Tmingc}
\end{figure}

We simulated the skymap for different observation durations and estimated the minimum time needed to properly reconstruct the morphology of the source. This is defined as the minimum time such that the relative error on the reconstructed parameter (both center of gravity and size) is reduced to at least $1\%$ and its mean value is not more than 3 sigma away from its simulated value. We also required a minimum of 10 signal photons on the source region. In Fig. \ref{fig:Tmingc} the minimum time needed to reconstruct the center of gravity is shown as a function of the flux strength in four energy intervals\footnote{Note that for the shorter observation times the results should be taken with caution since the analytical parameterizations shown in \S\ref{cap:detector} are obtained from the CTA simulations optimized for 50 hours observation time.}. 

\begin{figure}[htbp]
     \begin{center}
     \subfigure{%
     \includegraphics[width=.93\columnwidth]{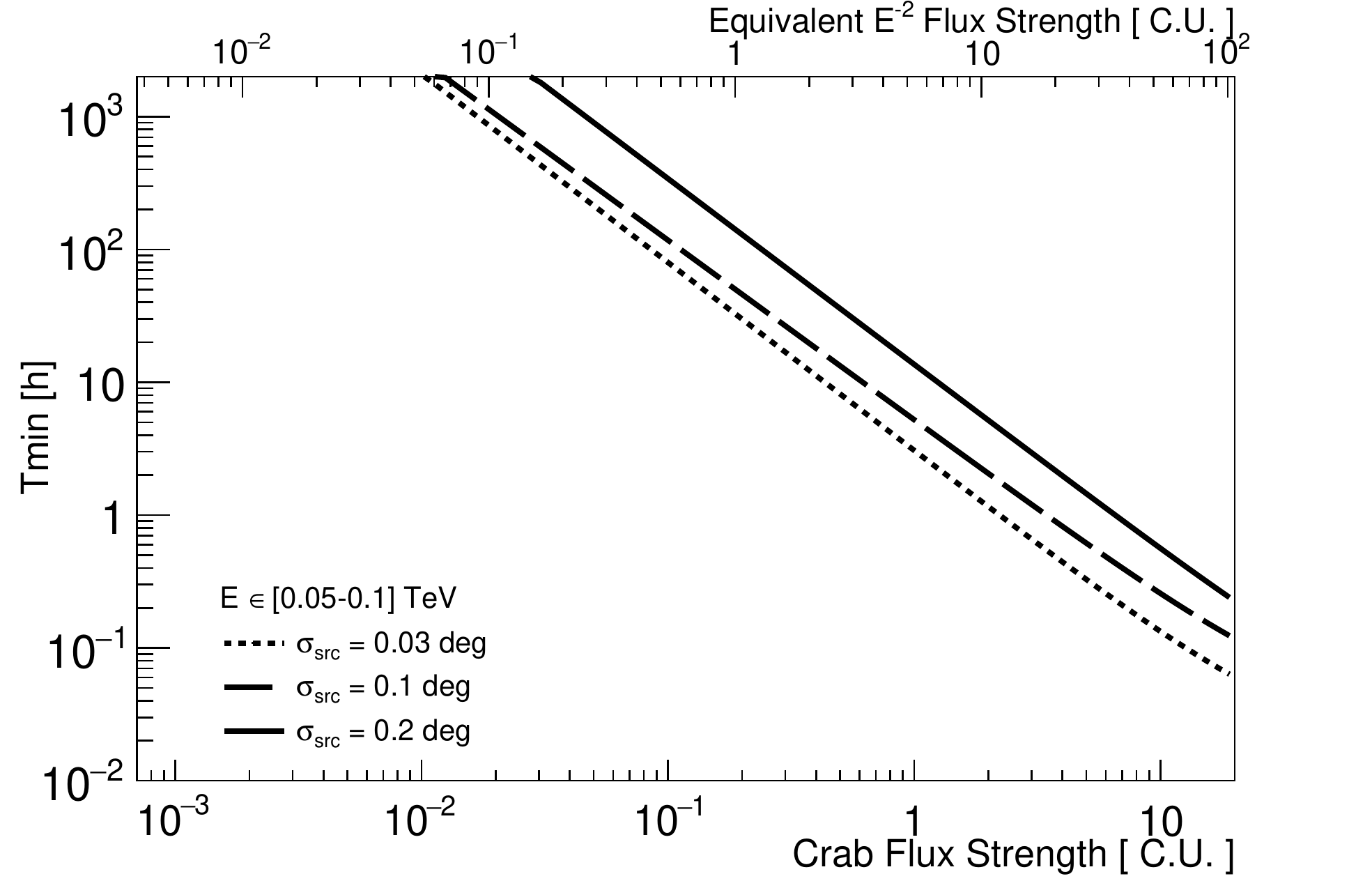}
      } \\
      \subfigure{%
      \includegraphics[width=.93\columnwidth]{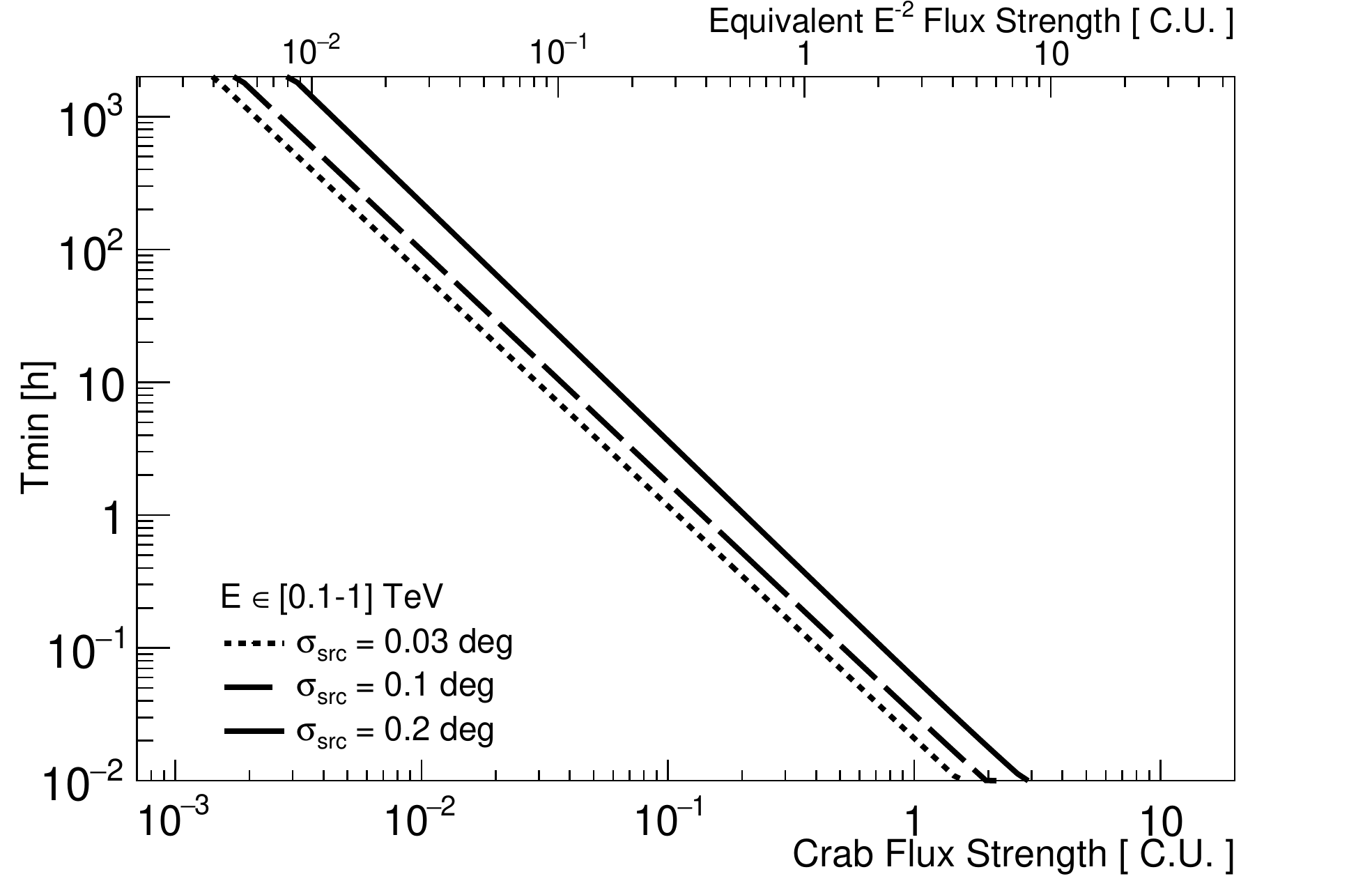}
      } \\
      \subfigure{%
      \includegraphics[width=.93\columnwidth]{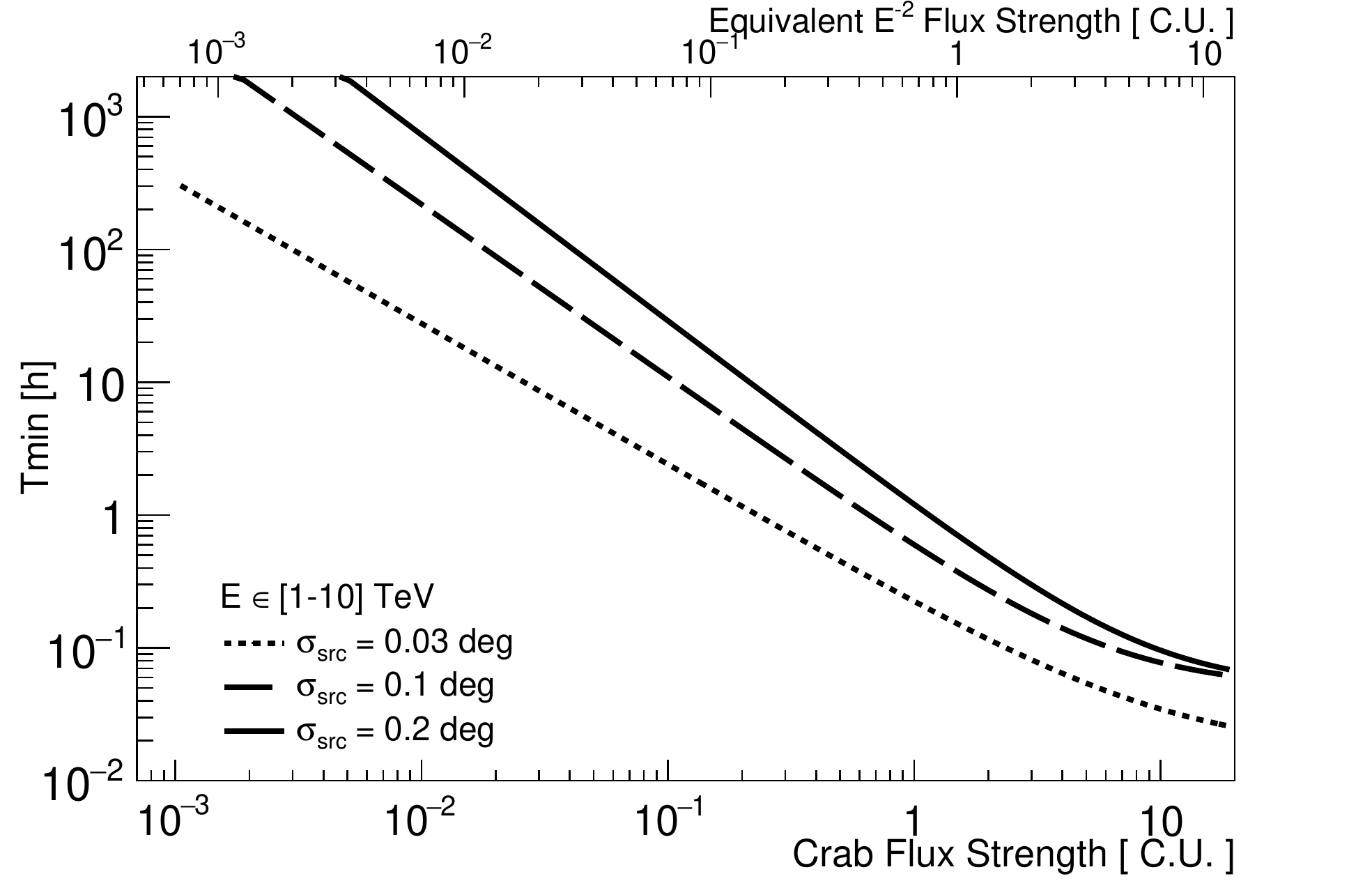}
      } \\
      \subfigure{%
      \includegraphics[width=.93\columnwidth]{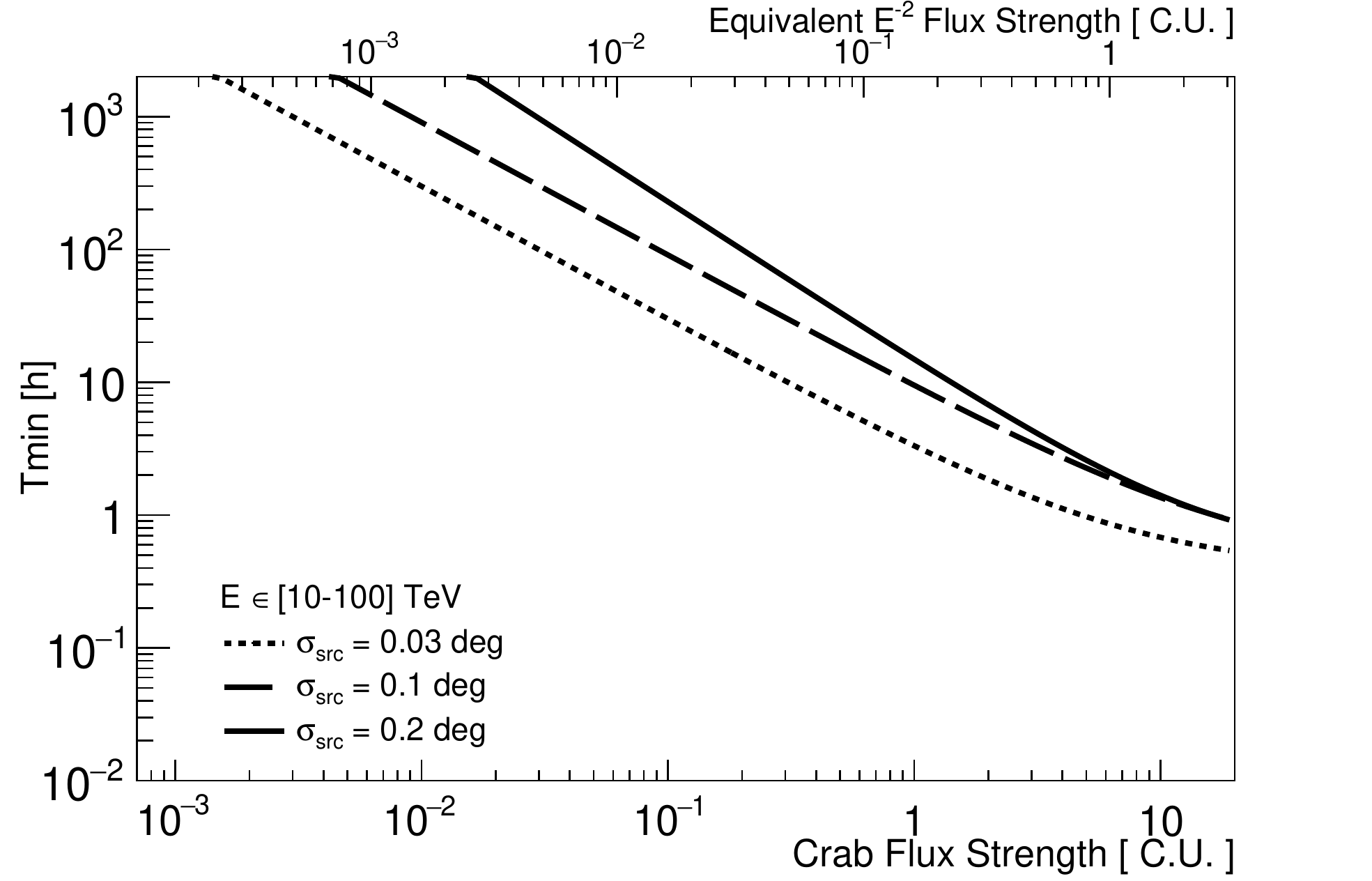}
      }
    \end{center}
\caption{Minimum time needed to reconstruct the size of the source as a function of the gamma-ray flux in the case of Gaussian PSF response and for four energy intervals: $[0.05-0.1]$\,TeV, $[0.1-1]$\,TeV, $[1-10]$\,TeV and $[10-100]$\,TeV. }
\label{fig:Tminsigma}
\end{figure}

One can see that the shortest minimum time is required in the energy 100\,GeV-1\,TeV and 1-10\,TeV bands, where the combination of best performance and reasonable statistics are found. At very low (50-100\,GeV) and very high (10-100\,TeV) energies the required time is significantly increased because of the poor performance and low photon statistics, respectively.

One should also note that the minimum time required to reconstruct the source position is always larger for extended sources than for point-like sources. The reason being that the flux is normalized to the source size and therefore the photon density is lower than in the case of point sources.

The same tendency is observed for the minimum time needed to estimate the source size. The results are shown in Fig. \ref{fig:Tminsigma}.

\subsection{Reconstruction of morphological parameters for non-Gaussian PSF response}  
\label{cap:singlesourcePSFtails}

\begin{figure*}[!ht]
\begin{minipage}[b]{8.5cm}
\centering
\includegraphics[width=8.5cm]{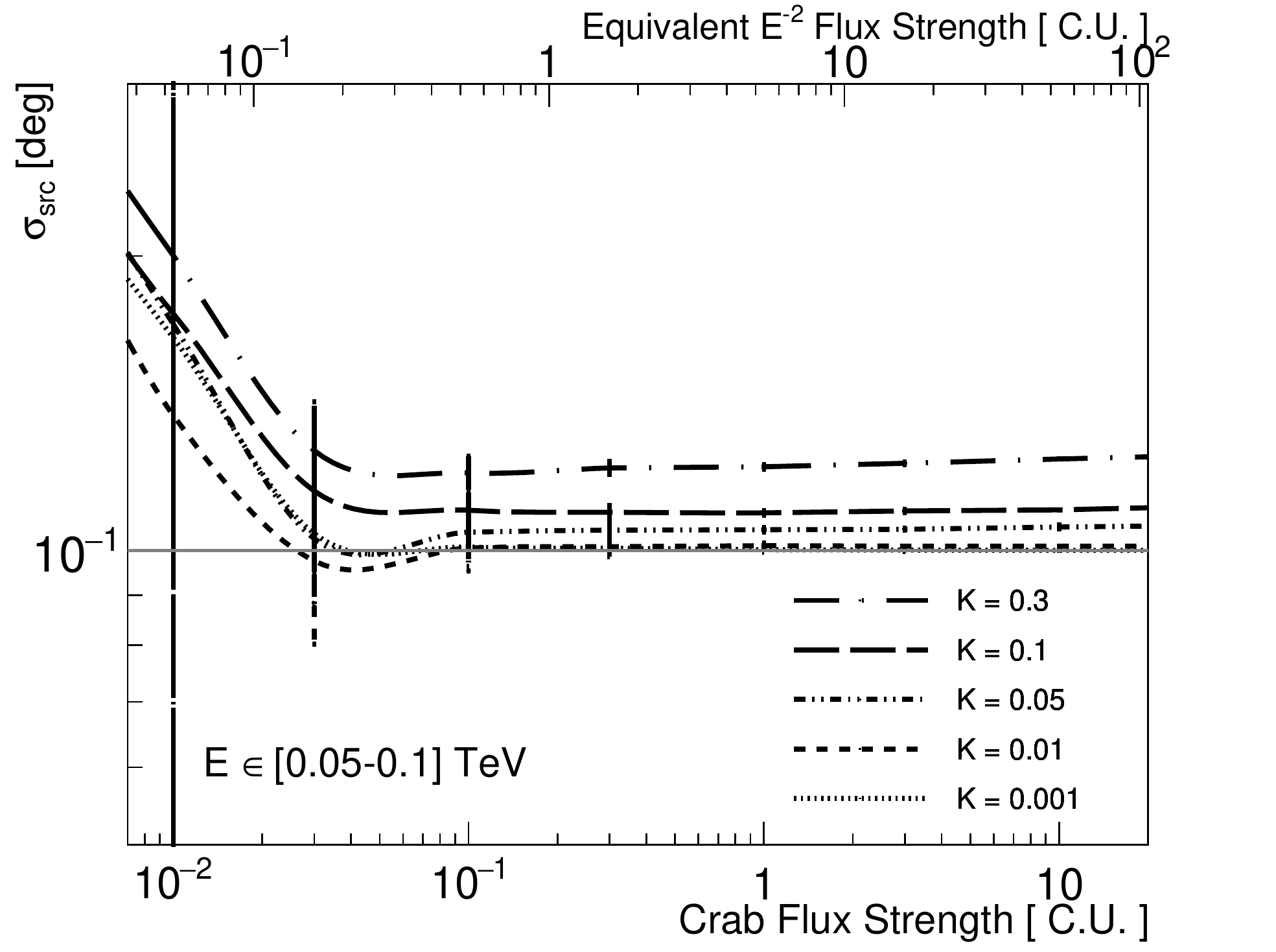}
\vspace{2mm}
\end{minipage}
\begin{minipage}[b]{8.5cm}
\centering
\includegraphics[width=8.5cm]{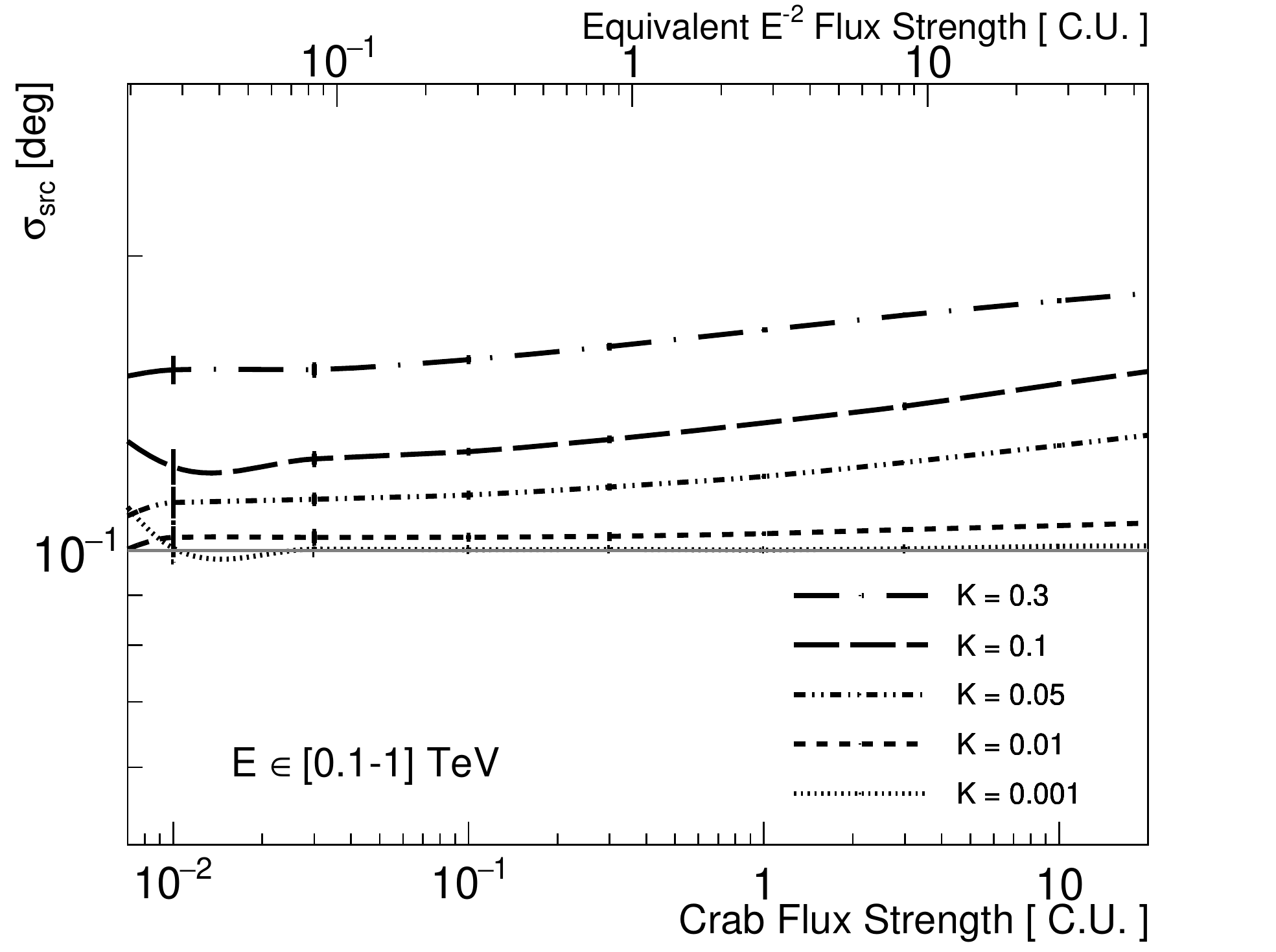}
\vspace{2mm}tre
\end{minipage}
\begin{minipage}[b]{8.5cm}
\centering
\includegraphics[width=8.5cm]{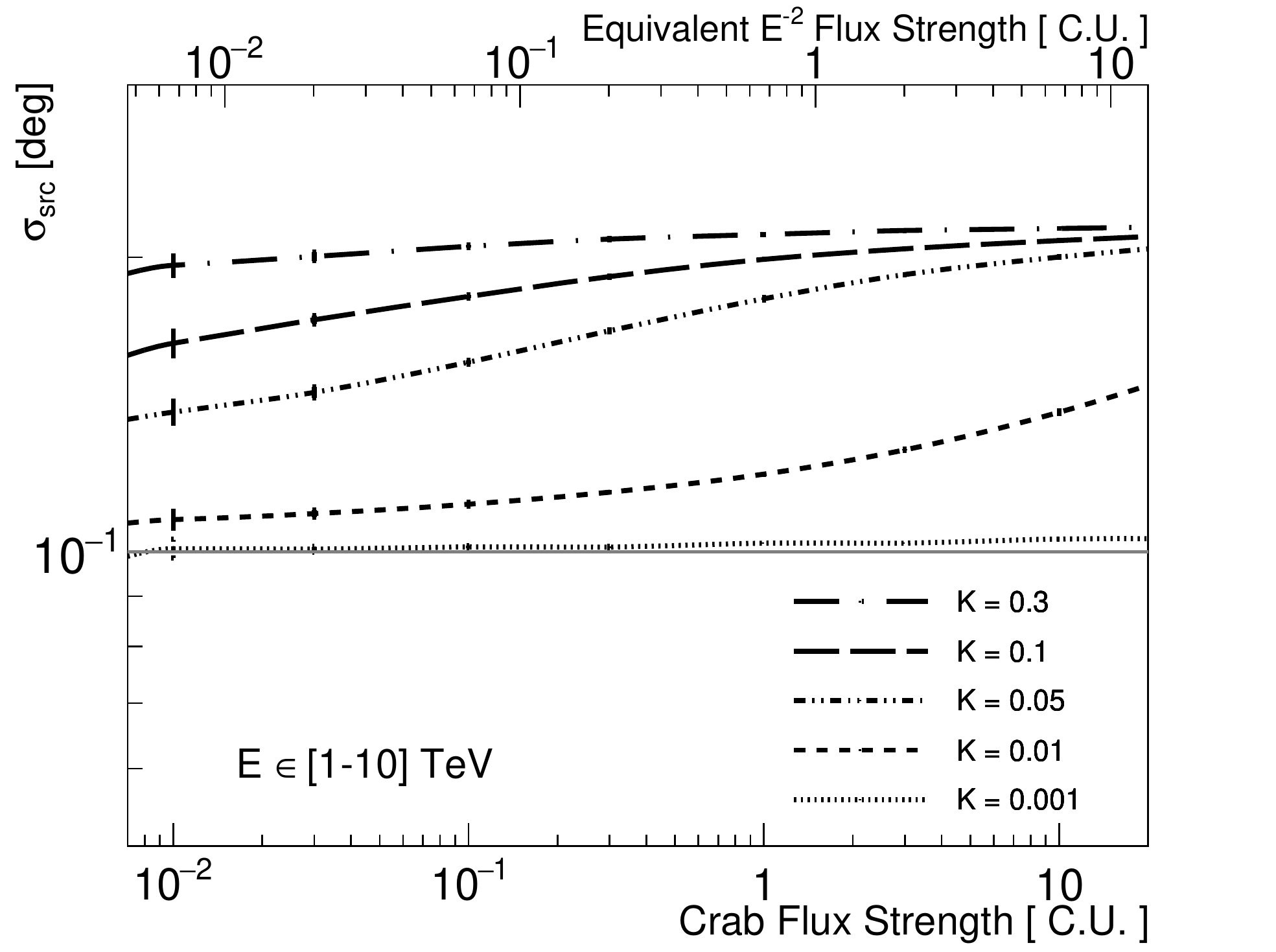}
\end{minipage}
\begin{minipage}[b]{8.5cm}
\centering
\includegraphics[width=8.5cm]{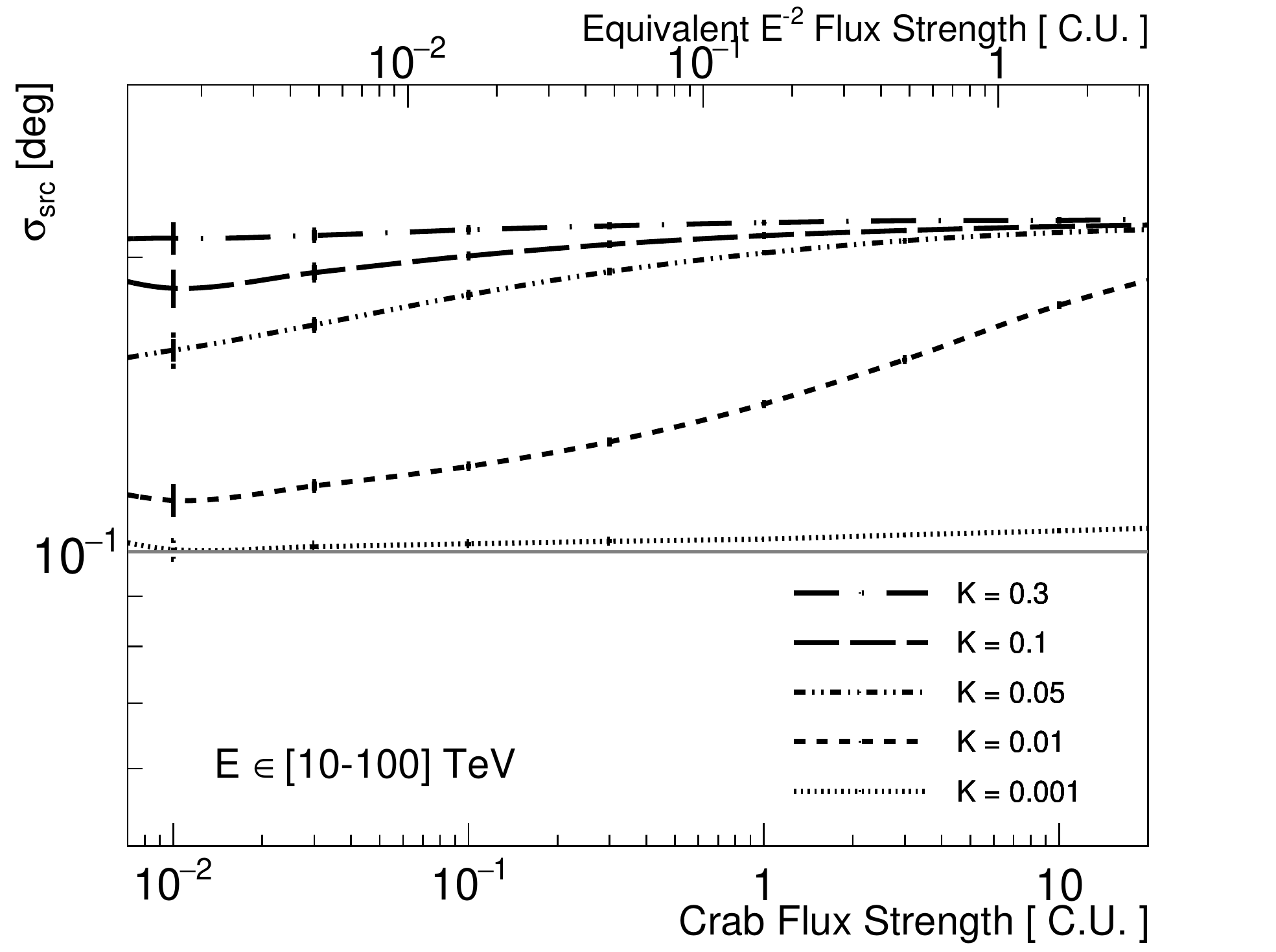}
\end{minipage}
\caption{Reconstructed size of the source as a function of the flux strength for different values of the parameter $K$ which defines the PSF shape composed with superposition of two Gaussians. The input value for the source size is indicated as a solid gray line, $\sigma_{src}=0.1$\,deg. The observation time is 50 hours. Four energy intervals are shown: $[0.05-0.1]$\,TeV, $[0.1-1]$\,TeV, $[1-10]$\,TeV and $[10-100]$\,TeV.}
\label{fig:tails_src01}
\end{figure*}

In this subsection we repeated the simulations described in the previous subsection using a non-Gaussian PSF shape (see Eq. \ref{eq:PSFtails}). The gamma-ray source is reconstructed using Eq. [\ref{eq:recofunc}] to investigate the effect of tails on the morphological reconstruction when these are not properly evaluated. The results are summarized in Fig. \ref{fig:tails_src01}. Each panel corresponds to the reconstructed source size obtained for the four energy intervals defined in \S\ref{cap:detector}. The reconstructed value is shown for different flux levels and different values of the parameter $K$ in Eq. \ref{eq:PSFtails}. The grey solid line at $0.1$\,deg indicates the size value used in the simulation.  

In the lowest energy domain ($[0.05, 0.1]$\,TeV), where the background dominates, the results for sources with fluxes $\le 0.1$\,C. U. are strongly affected by fluctuations. For brighter sources a proper estimation of the size can be done when the ratio $K$ is in the $0.01$ to $0.001$ range. The value of the PSF ($\sigma_{PSF}=0.147$\,deg) at these energies, comparable to the $\sigma_{PSFtails}=0.2$\,deg, results in a relatively small effect of the tails for small $K$ values. As expected, when $K$ is very large ($K\sim0.3$), the PSF cannot be described by a simple single Gaussian function and the effect of the wider distribution prevents a proper reconstruction of the real size of the source.

The value of $\sigma_{PSF}$ improves with energy (see Fig. \ref{fig:psf}), resulting in a more dramatic effect when adding $\sigma_{PSFtails}$ to the description of the PSF shape. The results in Fig. \ref{fig:tails_src01} shows that the larger the ratio $K$, the more significant the deviation from the expected value. This is true only when the tail contributions amounts to at least 5$\%$ of the peak value.

\section{Detection sensitivity for multiple sources in the same FoV} 
\label{multiplesources}

The study of two objects side by side has been performed. A first point-like gamma-ray source has been placed in the proximity of a second object. Hereafter we refer to the first source as {\itshape test source}, being the target of the observation, and to the other one as {\itshape background source}, since the gamma photons emitted by this nearby object represent an additional source of background for the \emph{test source}.

\subsection{Two neighbor sources} 
\label{}

The Crab-like spectrum defined by Eq. [\ref{eq:CrabFlux}] has been used both for the \emph{test} and the \emph{background source}. The shape of the two objects has been simulated according to the Gaussian distribution in Eq. [\ref{eq:gaussianshape}], characterized by the angular sizes $\sigma_{testSrc}$ and $\sigma_{bkgSrc}$ for which three different cases have been considered:
\begin{itemize}
	\item $\sigma_{testSrc}=0.03$\,deg and $\sigma_{bkgSrc}=0.03$\,deg;
	\item $\sigma_{testSrc}=0.03$\,deg and $\sigma_{bkgSrc}=0.1$\,deg;
	\item $\sigma_{testSrc}=0.03$\,deg and $\sigma_{bkgSrc}=0.2$\,deg.
\end{itemize}	

\begin{figure}[!ht]
\begin{minipage}[b]{8.5cm}
\centering
\includegraphics[width=8.5cm]{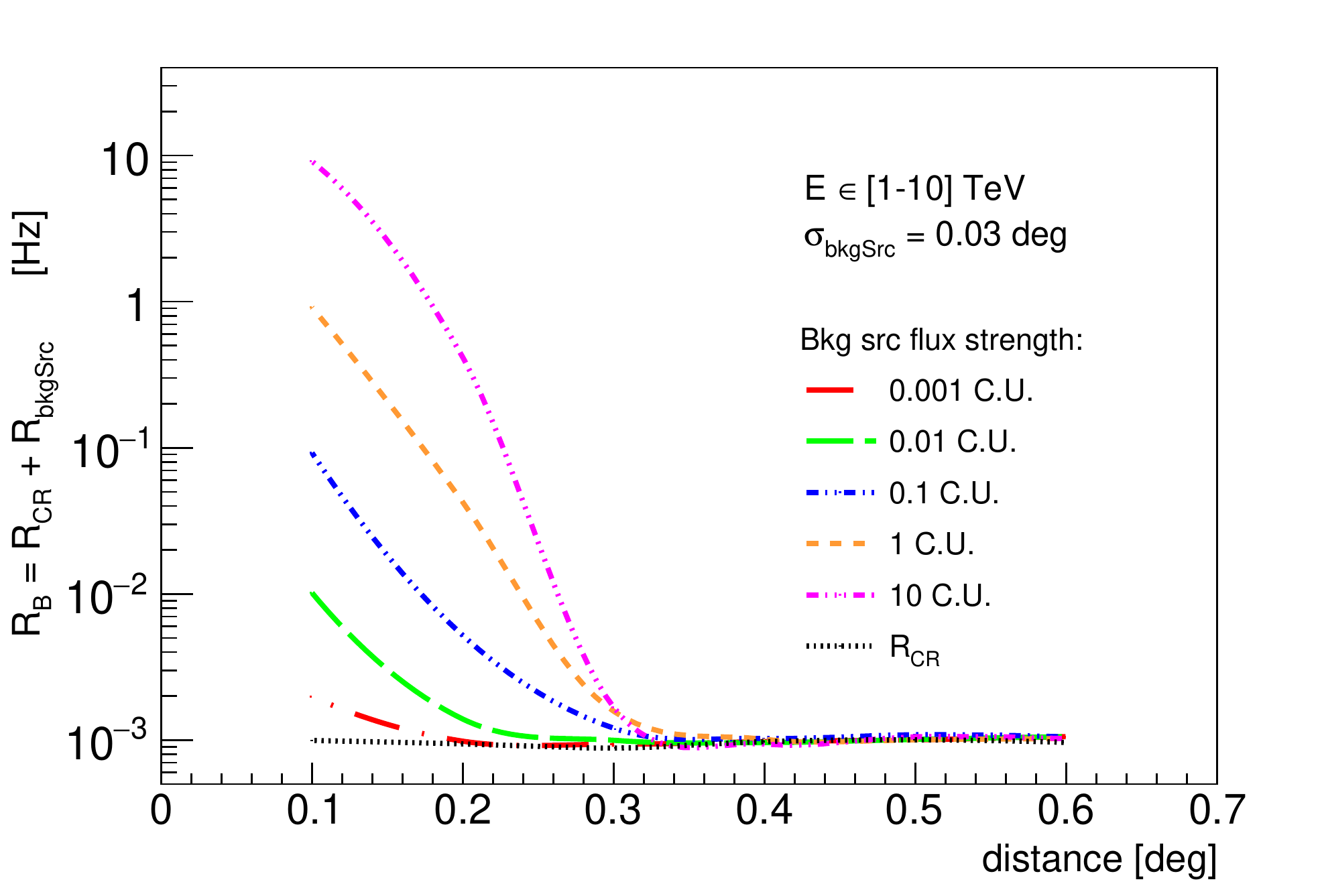}
\end{minipage}
\begin{minipage}[b]{8.5cm}
\centering
\includegraphics[width=8.5cm]{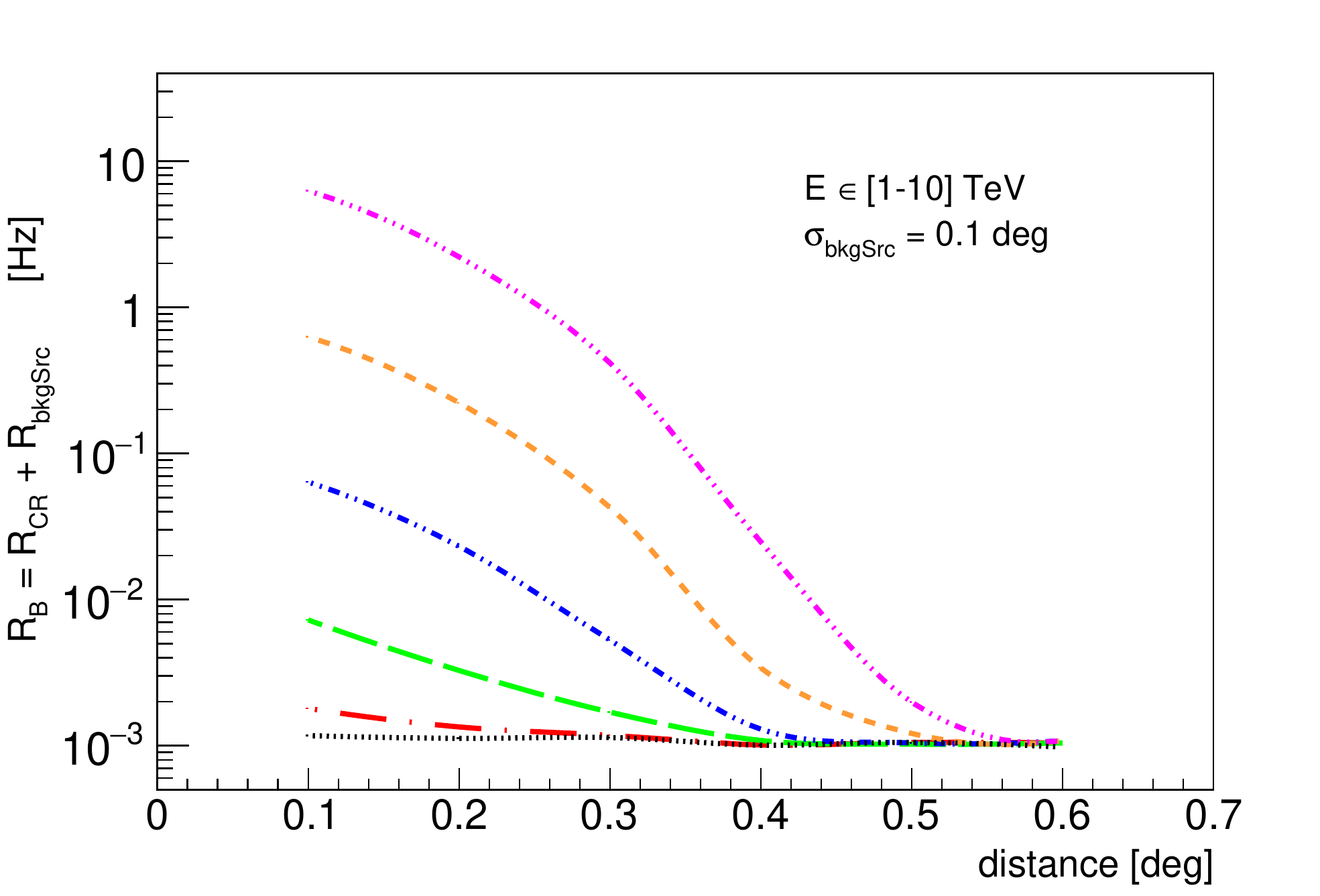}
\end{minipage}
\begin{minipage}[b]{8.5cm}
\centering
\includegraphics[width=8.5cm]{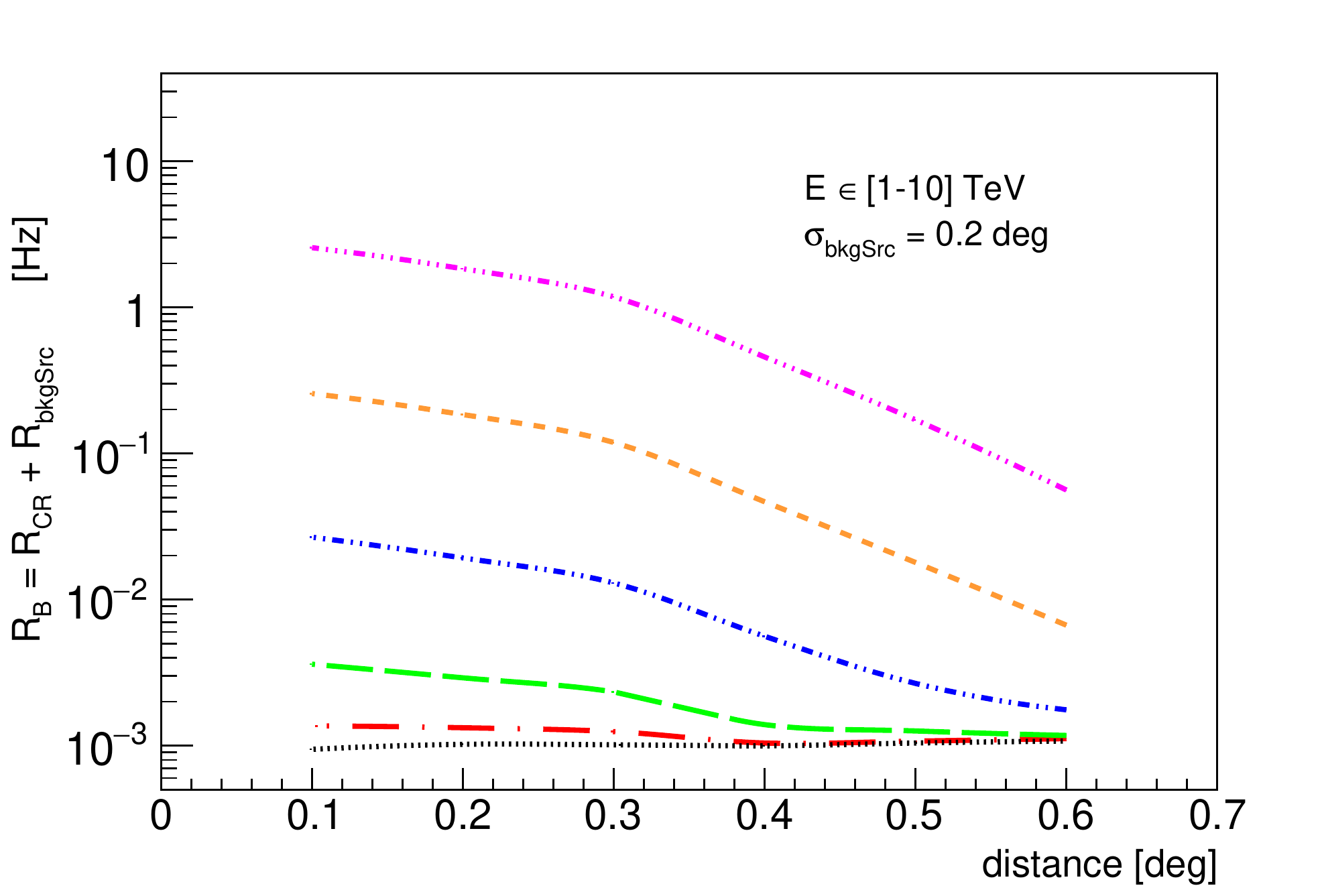}
\end{minipage}
\caption{Total background rate as a function of the distance between the \emph{test source} and the nearby gamma-ray \emph{background source}, assuming a Gaussian PSF. The total background rate $R_B$ is defined as the sum of the cosmic ray background, $R_{CR}$ (black dotted line), and the additional noise due to the photons coming from the \emph{background source}, $R_{bkgSrc}$. The calculations in the energy interval from 1\,TeV to 10\,TeV are conducted for different flux strengths and angular sizes of the \emph{background source} indicated on the figures.}
\label{fig:bkgrateVSd}
\end{figure}

\begin{figure}[!ht]
\begin{minipage}[b]{8.5cm}
\centering
\includegraphics[width=8.5cm]{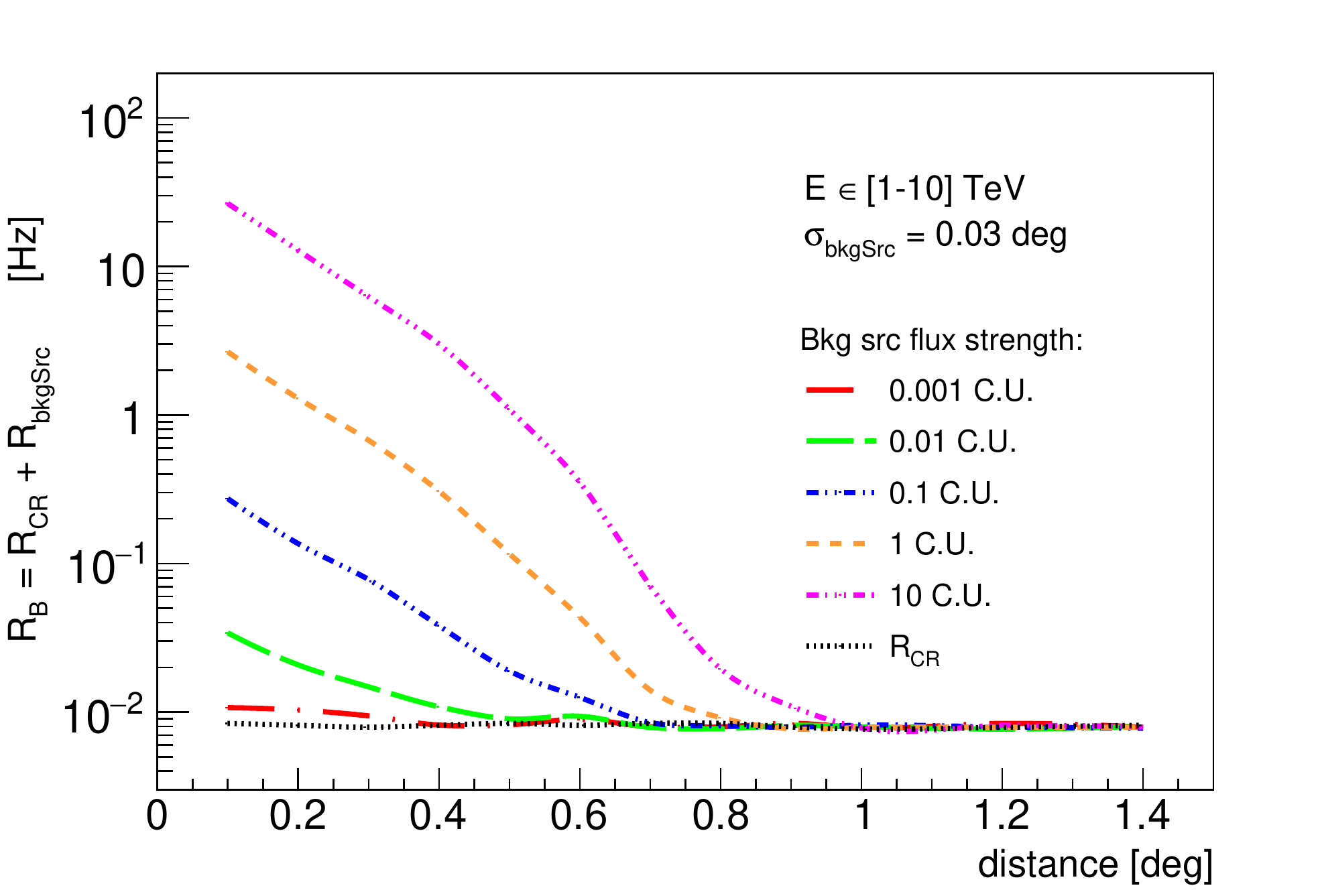}
\end{minipage}
\begin{minipage}[b]{8.5cm}
\centering
\includegraphics[width=8.5cm]{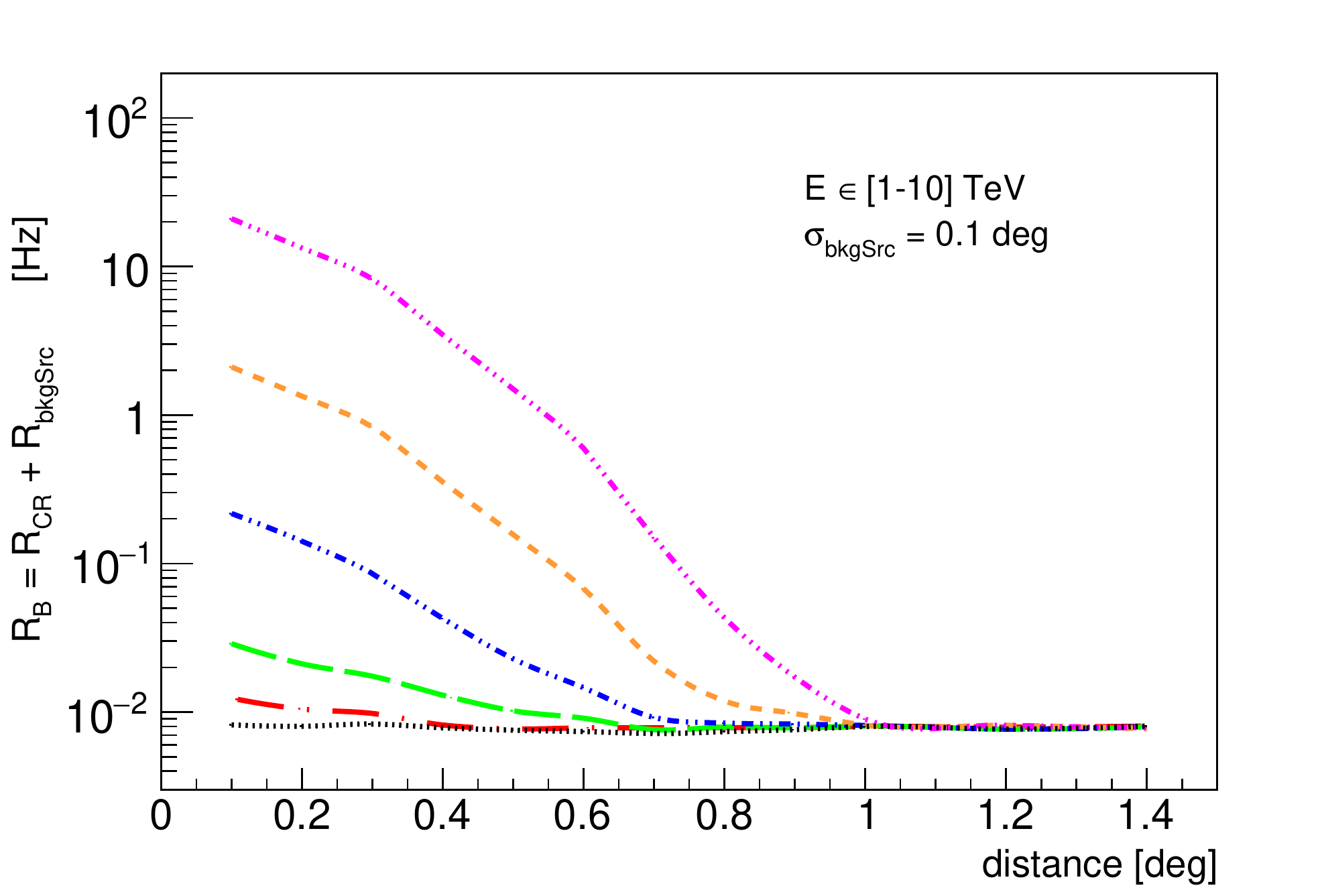}
\end{minipage}
\begin{minipage}[b]{8.5cm}
\centering
\includegraphics[width=8.5cm]{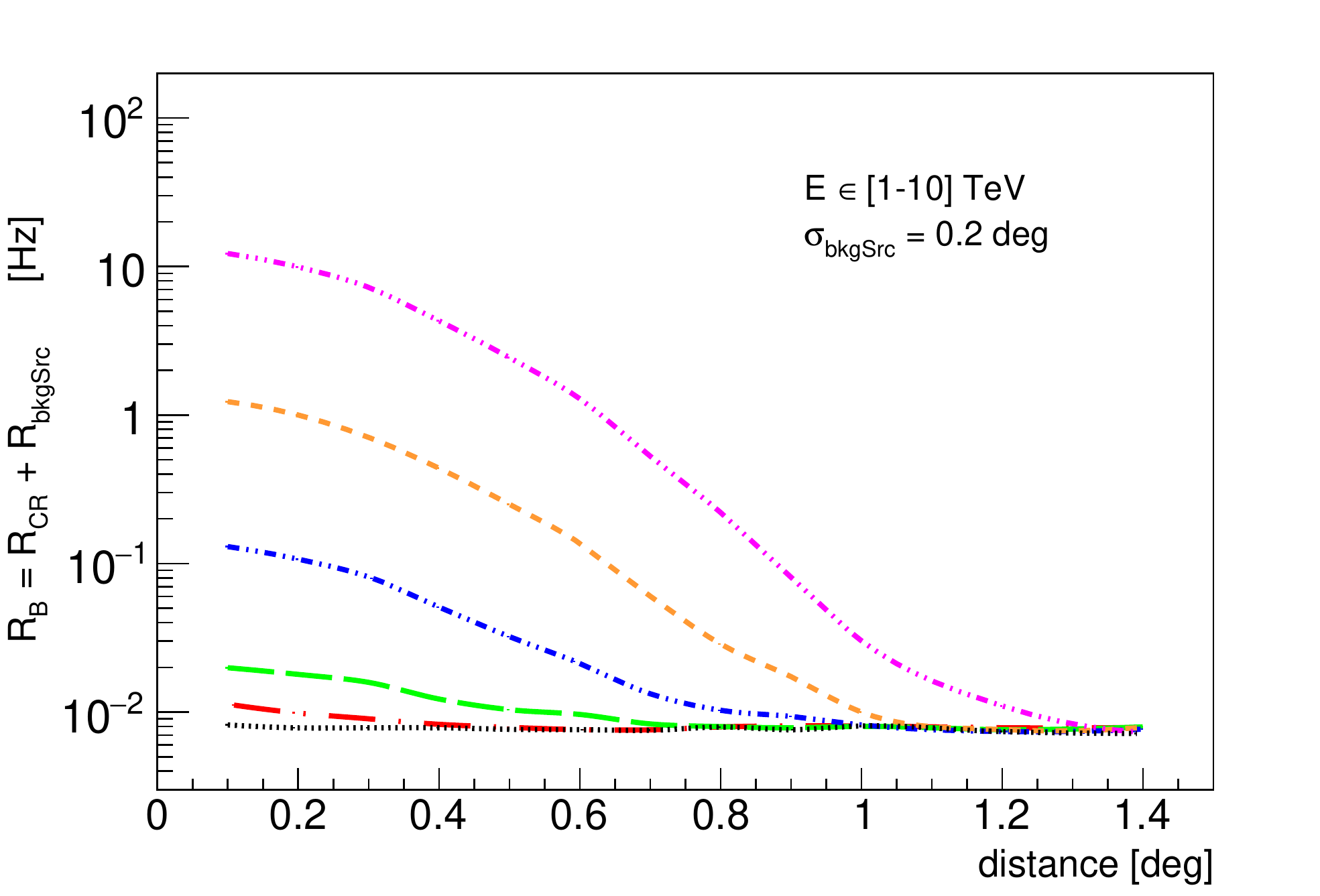}
\end{minipage}
\caption{Same as Fig. \ref{fig:bkgrateVSd} but assuming a non-Gaussian PSF with tails described by Eq. [\ref{eq:PSFtails}] for $K=0.3$.}
\label{fig:bkgrateVSd_tails}
\end{figure}

\subsubsection{Event rate} 
\label{}

Gamma-ray photons from the \emph{background source} reaching the region of the observation limit the ability to resolve the main object. Therefore, the closer the \emph{background source}, the higher the total amount of background. This effect is clearly seen in Fig. \ref{fig:bkgrateVSd}, where the total background rate $R_B = R_{CR} + R_{bkgSrc}$ is shown as a function of the distance between the two sources in the case of a Gaussian PSF. $R_{bkgSrc}$ represents the additional noise due to the photons coming from the \emph{background source}, while $R_{CR}$ is the standard background rate due to the cosmic rays. Comparing different hypothesis for the \emph{background source} size, we found that extended sources ($\sigma_{bkgSrc} > \sigma_{PSF}$) contaminate the cosmic ray background rate up to $0.6$\,deg separation from the \emph{test source}. On the contrary, when the nearby \emph{background source} is point-like, this effect on the background rate disappears as soon as the distance between the two objects is larger than $0.3$\,deg, irrespective of the flux strength of the companion.

Similar calculations have been conducted under the assumption of a non-Gaussian PSF: the results are shown in Fig. \ref{fig:bkgrateVSd_tails}. Apparently, the tails contribute significantly to the background, especially at large distances from the gamma-ray \emph{background source}. One can see that for the parameters used in these calculations, a significant contribution from the \emph{background source} can extend up to $1$\,degree, i.e. an order of magnitude larger than the PSF.  A comparison of Fig. \ref{fig:bkgrateVSd} and Fig. \ref{fig:bkgrateVSd_tails} shows that this effect is apparently caused by the tails of the PSF.

\begin{figure*}[p]
\begin{minipage}[b]{\textwidth}
\centering
\includegraphics[width=\textwidth]{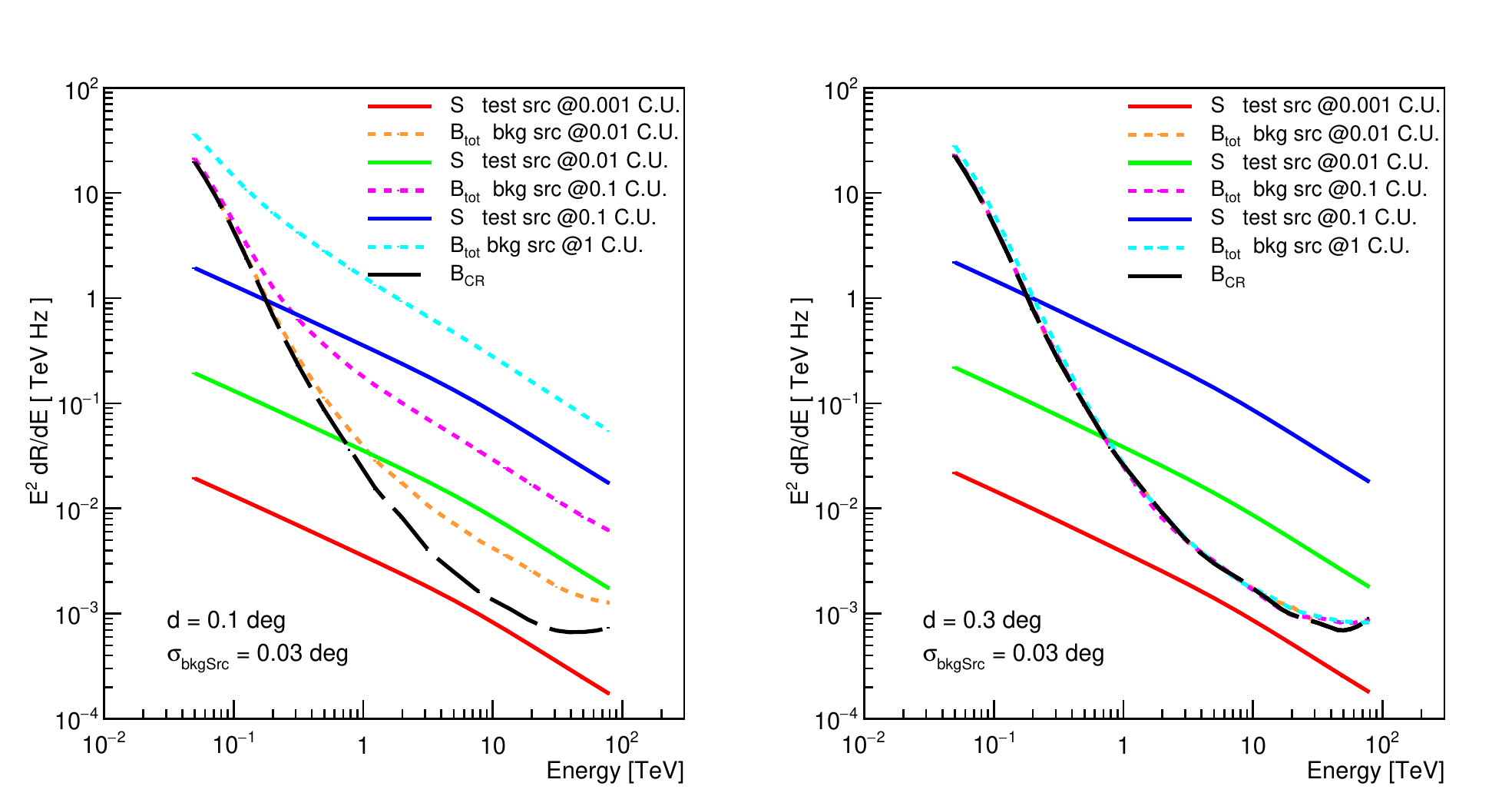}
\end{minipage}
\begin{minipage}[b]{\textwidth}
\centering
\includegraphics[width=\textwidth]{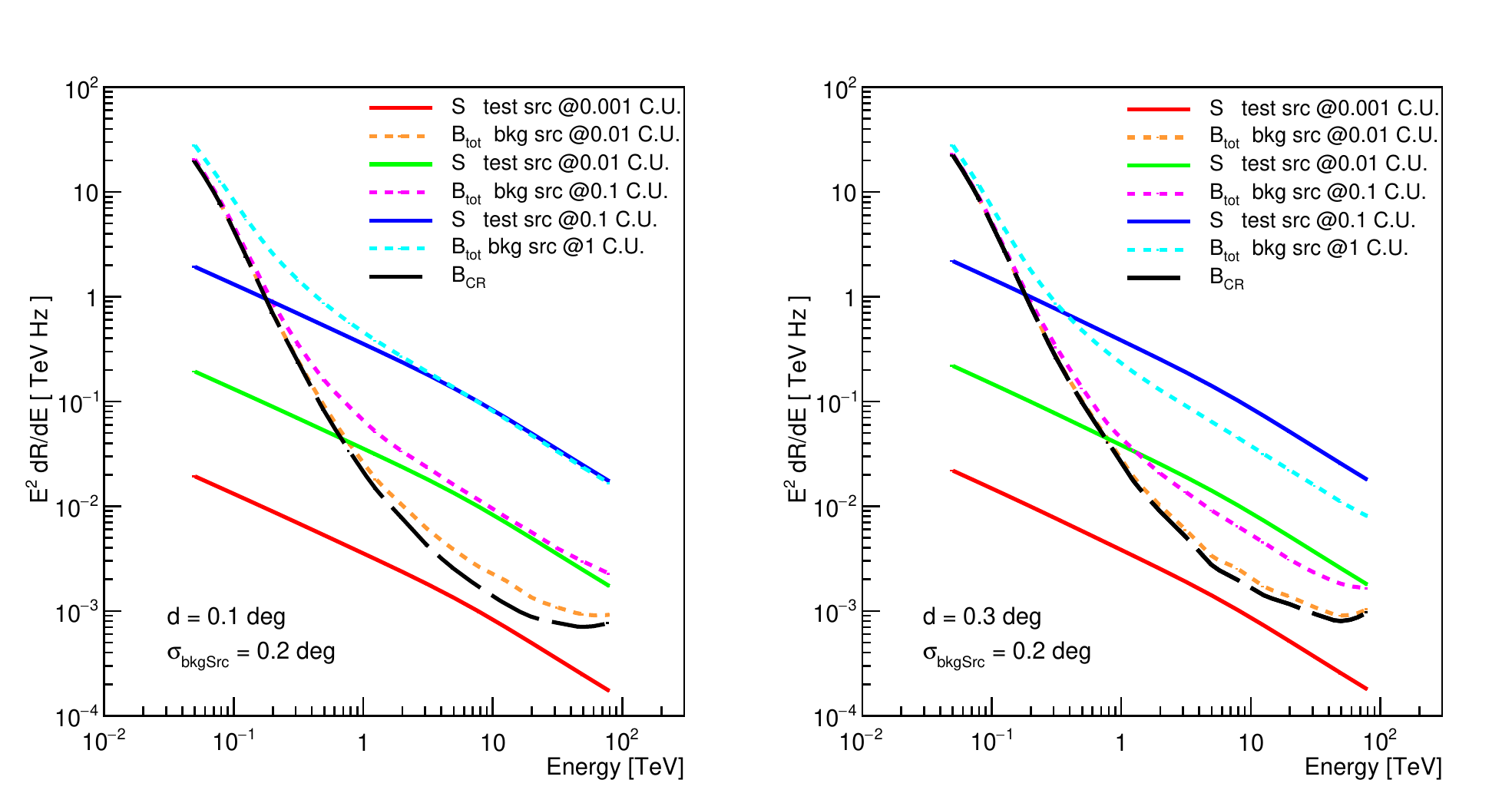}
\end{minipage}
\caption{Differential event rates as a function of the energy, assuming a Gaussian PSF. Solid lines are for the gamma-ray signal from the \emph{test source}. Dashed lines are for the total background ($B_{tot}$) which consists of the background induced by cosmic rays $B_{CR}$ (black dashed curves) and background induced by gamma-rays from the neighbor gamma-ray \emph{background source}. The calculations have been performed for different sizes of the \emph{background source} (top panels for a point-like source, and the bottom panels for the second source with size $0.2$\,deg), for different distances between two sources (left panels:  $0.1$\,deg; right panels: $0.3$\,deg), and for different fluxes of the \emph{test} and \emph{background} gamma-ray sources (the numbers are indicated on figures).}
\label{fig:DetectionRate2src_20E}
\end{figure*}

\begin{figure*}[p]
\begin{minipage}[b]{\textwidth}
\centering
\includegraphics[width=\textwidth]{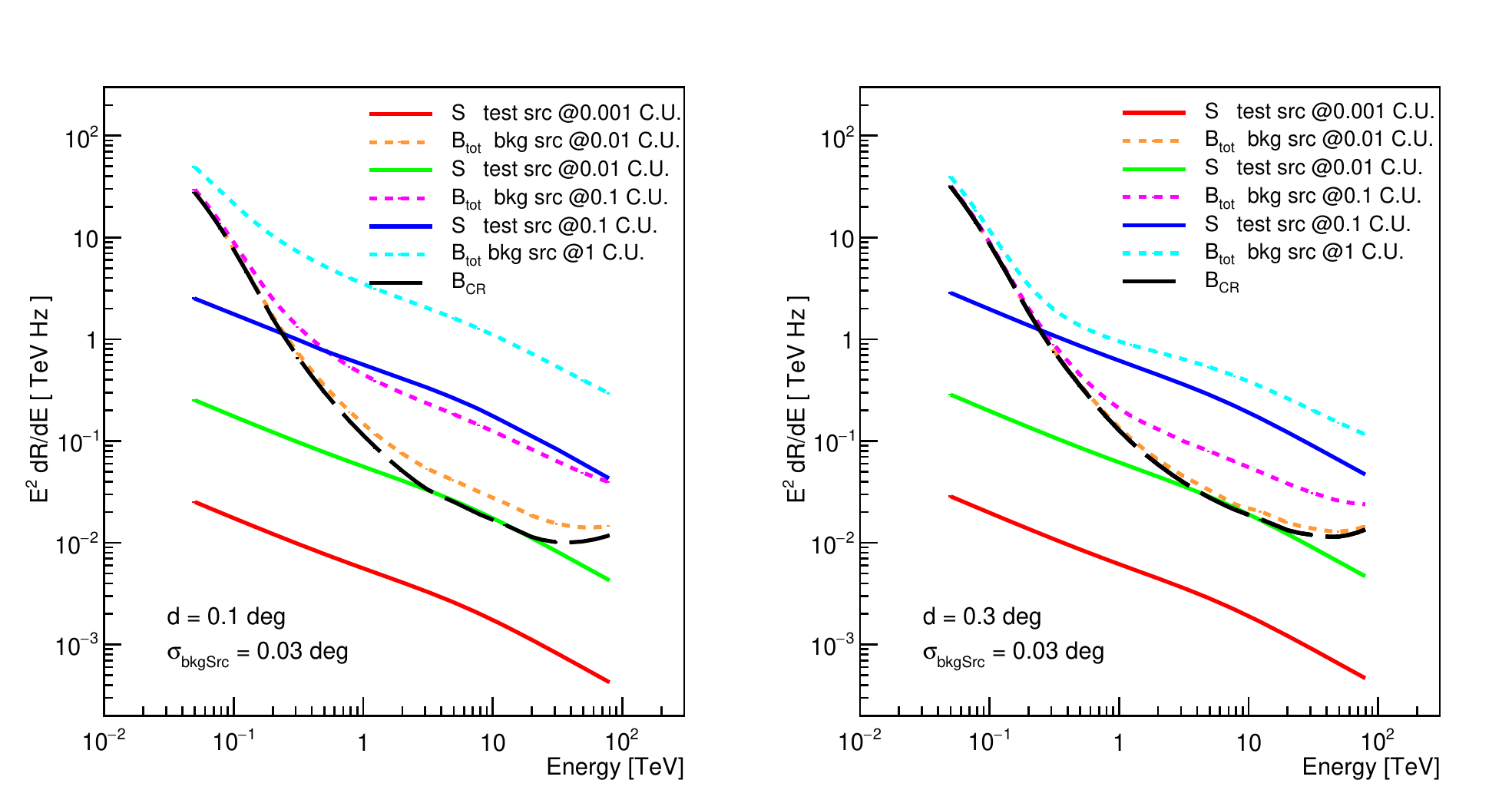}
\end{minipage}
\begin{minipage}[b]{\textwidth}
\centering
\includegraphics[width=\textwidth]{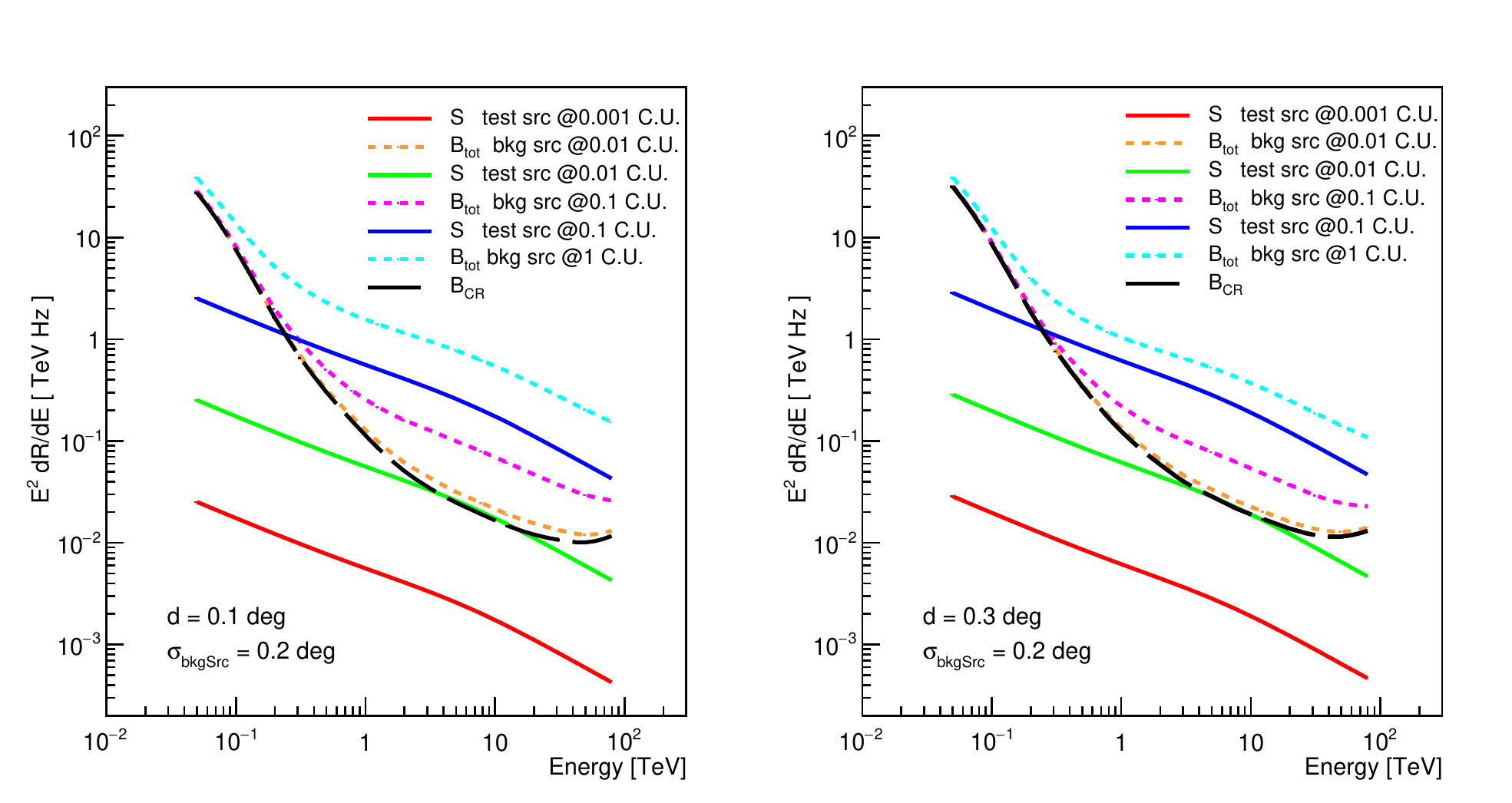}
\end{minipage}
\caption{Same as in Fig. \ref{fig:DetectionRate2src_20E} but assuming a non-Gaussian PSF with tails described by Eq. [\ref{eq:PSFtails}] with $K=0.3$.}
\label{fig:DetectionRate2src_20E_tails}
\end{figure*}

In Fig. \ref{fig:DetectionRate2src_20E} and \ref{fig:DetectionRate2src_20E_tails} we show the energy dependence of differential event rates for two different distances between the \emph{test} and the \emph{background} gamma-ray sources: $0.1$\,deg and $0.3$\,deg. The curves in Fig. \ref{fig:DetectionRate2src_20E} are obtained for the pure Gaussian distribution of the PSF, while Fig. \ref{fig:DetectionRate2src_20E_tails} corresponds to the non-Gaussian PSF with tails ($K=0.3$). The results of this section indicate that, in the presence of neighboring gamma-ray sources, the background for the \emph{test source} can significantly exceed (depending on the flux of the \emph{background source} and its distance from the \emph{test source}) the rate of background events induced by cosmic rays. This would result in a significant reduction of the sensitivity of observations, specifically in regions rich in gamma-ray sources.

\subsubsection{The sensitivity} 
\label{}

CTA foresees improving beyond the sensitivity of current ground-based high-energy gamma-ray instruments by an order of magnitude. This factor concerns the observations of a single point-like object, isolated from any other source. However, it is expected that such an improvement of sensitivity will dramatically increase the number of known gamma-ray sources. Correspondingly, the average distance between sources will also be reduced, especially in the Galactic plane. In this case the background for the test source caused by the surrounding gamma-ray sources can be comparable to, or even exceed, the background induced by cosmic rays. This implies that the sensitivity curves obtained under the assumption of only cosmic ray background may significantly overestimate the capability of the instrument to resolve weak sources.

Following the criteria used by the CTA Consortium to compute the differential sensitivity curve \citep{2013APh....43..171B}, we estimated the flux sensitivity for 50 hours of observation of an object with a spectrum given by Eq. [\ref{eq:CrabFlux}] placed close to a companion described by the same spectrum. The \emph{background source} flux is fixed to $0.1$\,C.U. Such a flux intensity represents a good compromise in the contest of our study, given the considerable number of objects in the Galactic plane having this flux strength (see e.g. ref. \cite{Carrigan:2013}) and the corresponding non-negligible effect on the \emph{test source} which might significantly constrain the observations. A discrete set of flux bins has been 

\FloatBarrier

\noindent
simulated for the \emph{test source}. This results in a sensitivity \emph{range} instead of a curve, defined by the edges of the flux bin in which the minimum detectable flux is reached. This minimum flux is obtained binning the energy spectrum from 50\,GeV to 100\,TeV in five independent logarithmic bins per decade of energy, as in \citep{2013APh....43..171B}. For each energy bin the requirements described in \S\ref{cap:CTAsensitivity} are applied.

\begin{figure}[!ht]
\centering
\includegraphics[width=8.5cm]{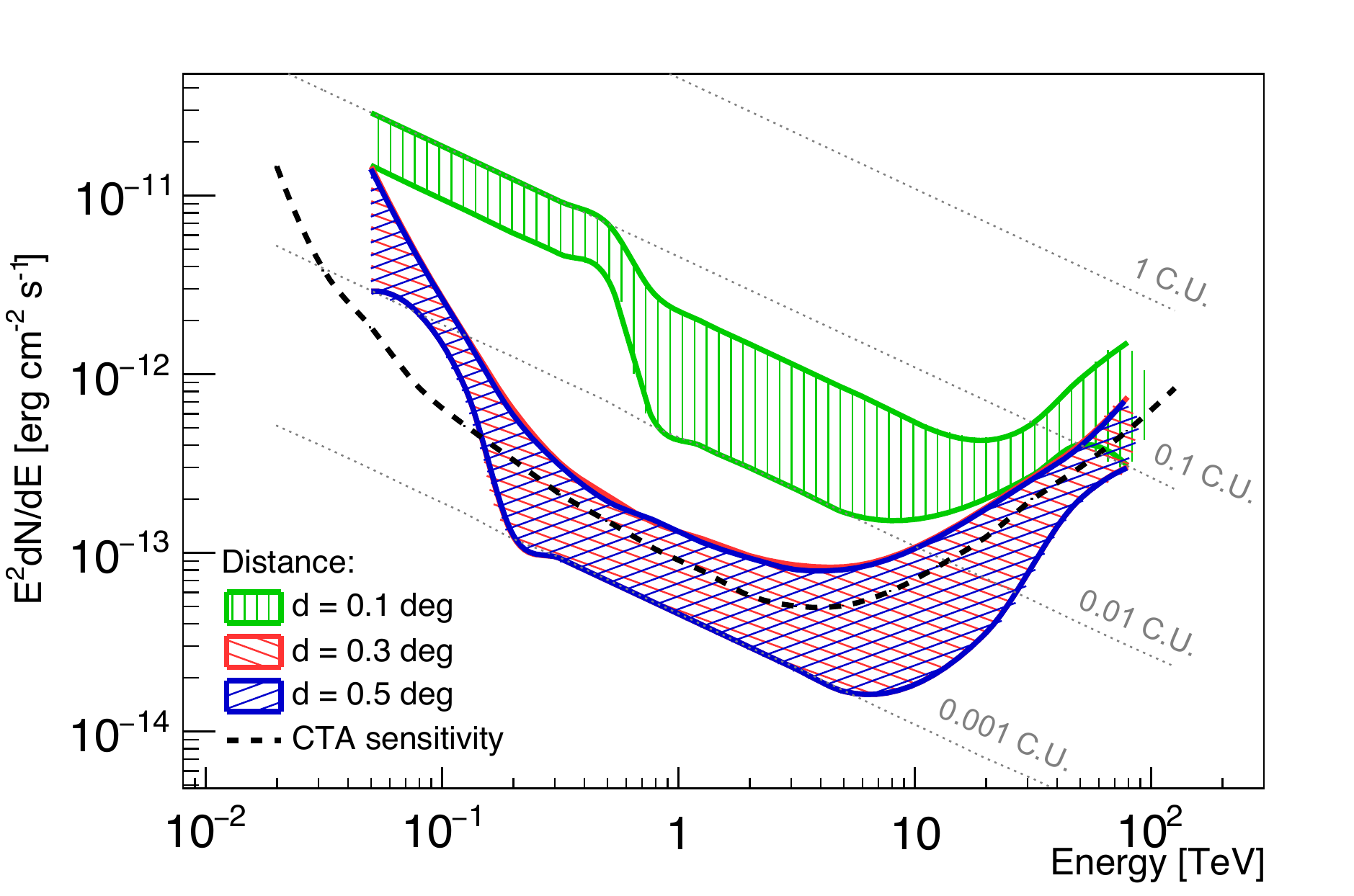}
\caption{Differential flux sensitivities for a point-like \emph{test source} corresponding to 50h observation time. The range of sensitivities are shown by shaded regions. The upper and lower edges of these regions indicate the uncertainties of simulations related to the discrete flux levels (see the text). The calculations are performed under the assumption of the existence of a neighbor point-like gamma-ray \emph{background source} of strength $0.1$\,Crab. The green, red and blue regions correspond to the location of the neighbor source at distances of $0.1$, $0.3$ and $0.5$\,degrees, correspondingly. In calculations we assumed a pure Gaussian distribution for the PSF. The publicly available sensitivity expected by the CTA for a point-like source is also reported in the figure (black dashed curve)\footref{CTAwebpage}.}
\label{fig:sens11}
\end{figure}

In Fig. \ref{fig:sens11}, we show the differential sensitivity (shaded area) to detect a point-like source when located in proximity to a second point-like source ($\sigma_{testSrc}=\sigma_{bkgSrc}=0.03$\,deg). The differential sensitivity is calculated for a distance between the two sources of $0.1$\,deg, $0.3$\,deg and $0.5$\,deg (in green, red and blue respectively). The dashed-black line shows the publicly available sensitivity curve for a point-like source obtained by the CTA Consortium\footref{CTAwebpage}. The latest was calculated under the assumption that the background is caused only by cosmic rays, therefore it does not depend on the location of the \emph{test source}. 

\begin{figure}[!ht]
\centering
\includegraphics[width=8.5cm]{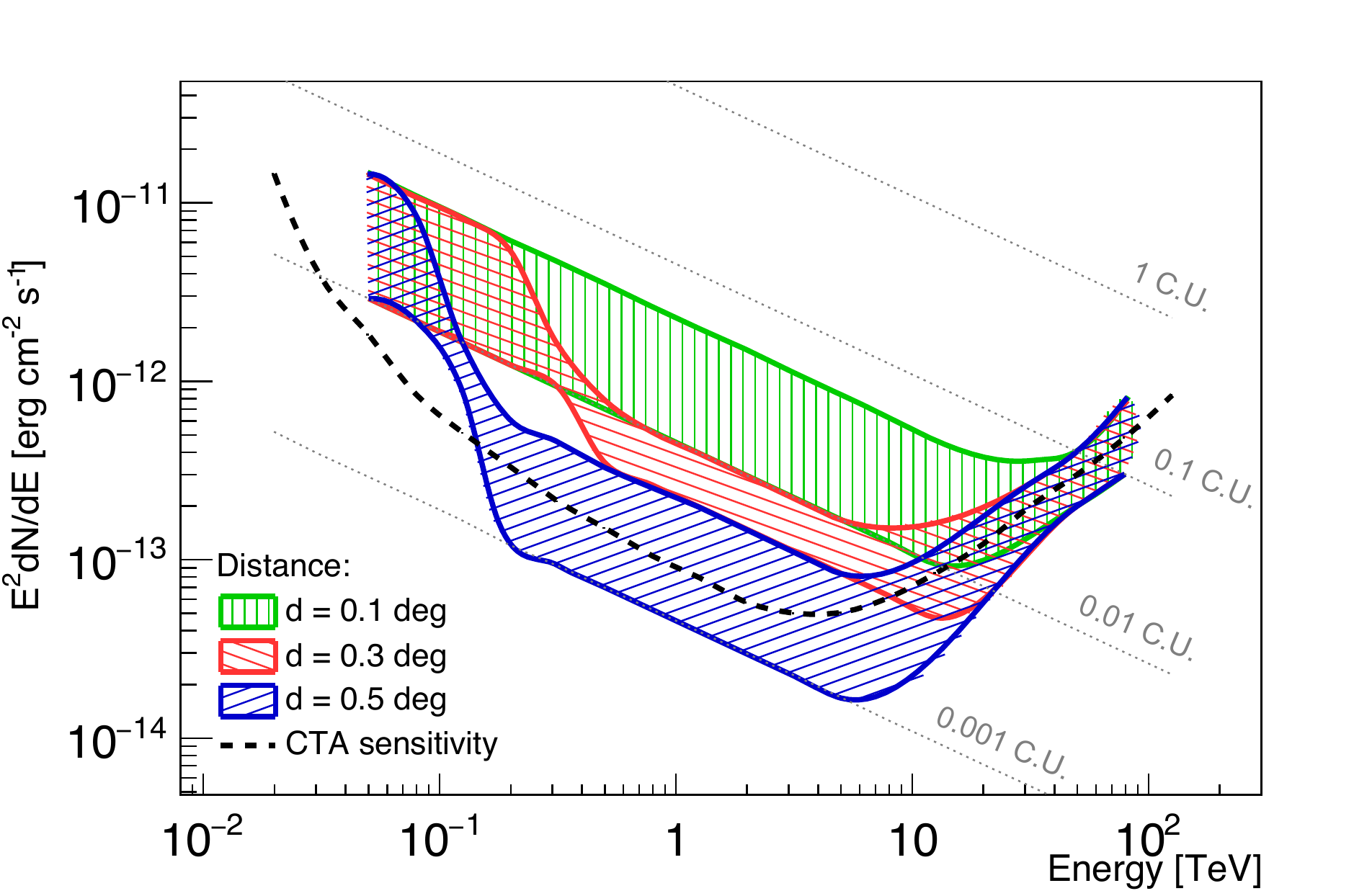}
\caption{Same as in Fig. \ref{fig:sens11} but calculated for an extended \emph{background gamma-ray source}.}
\label{fig:sens13}
\end{figure}

Fig. \ref{fig:sens11} shows that the sensitivity to detect a point-like source in the presence of even a moderately strong source of $0.1$\,C.U. can be significantly (up to an order of magnitude) worse than in the case of an isolated one. The effect cannot be ignored in environments densely populated by gamma-ray sources, as in the case of the Galactic plane. Since most of the Galactic sources are extended, in Fig. \ref{fig:sens13} we show the sensitivities in the regions around an extended gamma-ray source with angular size $0.2$\,deg, similar to the average size of the objects in the Galactic plane \cite{Carrigan:2013}. 

\begin{figure}[!ht]
\centering
\includegraphics[width=8.5cm]{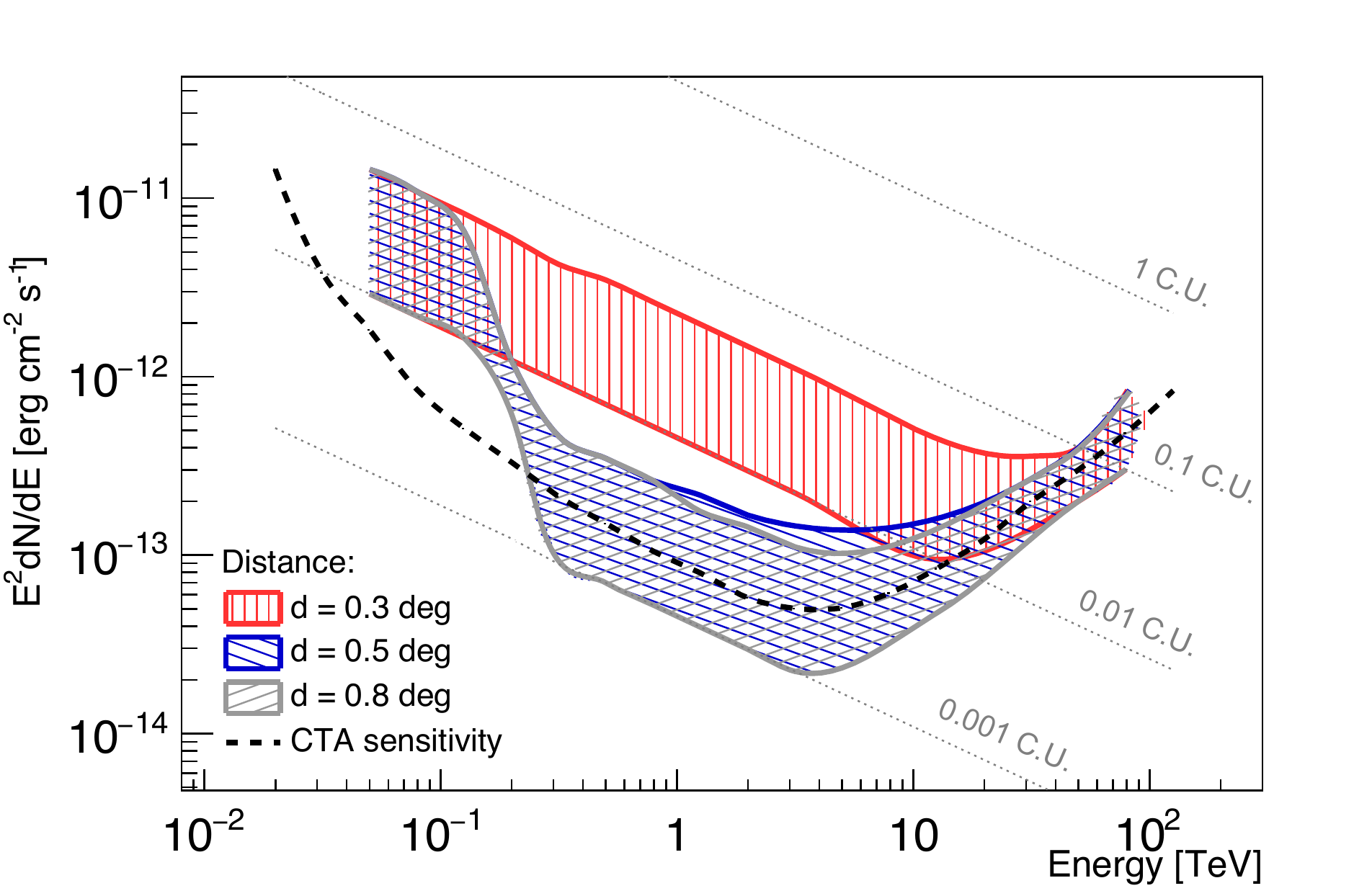}
\caption{Same as Fig. \ref{fig:sens11}, but for the case of non-Gaussian PSF with tails given by Eq. [\ref{eq:PSFtails}] with $K=0.3$.}
\label{fig:sens11_tails}
\end{figure}

\begin{figure}[!ht]
\centering
\includegraphics[width=8.5cm]{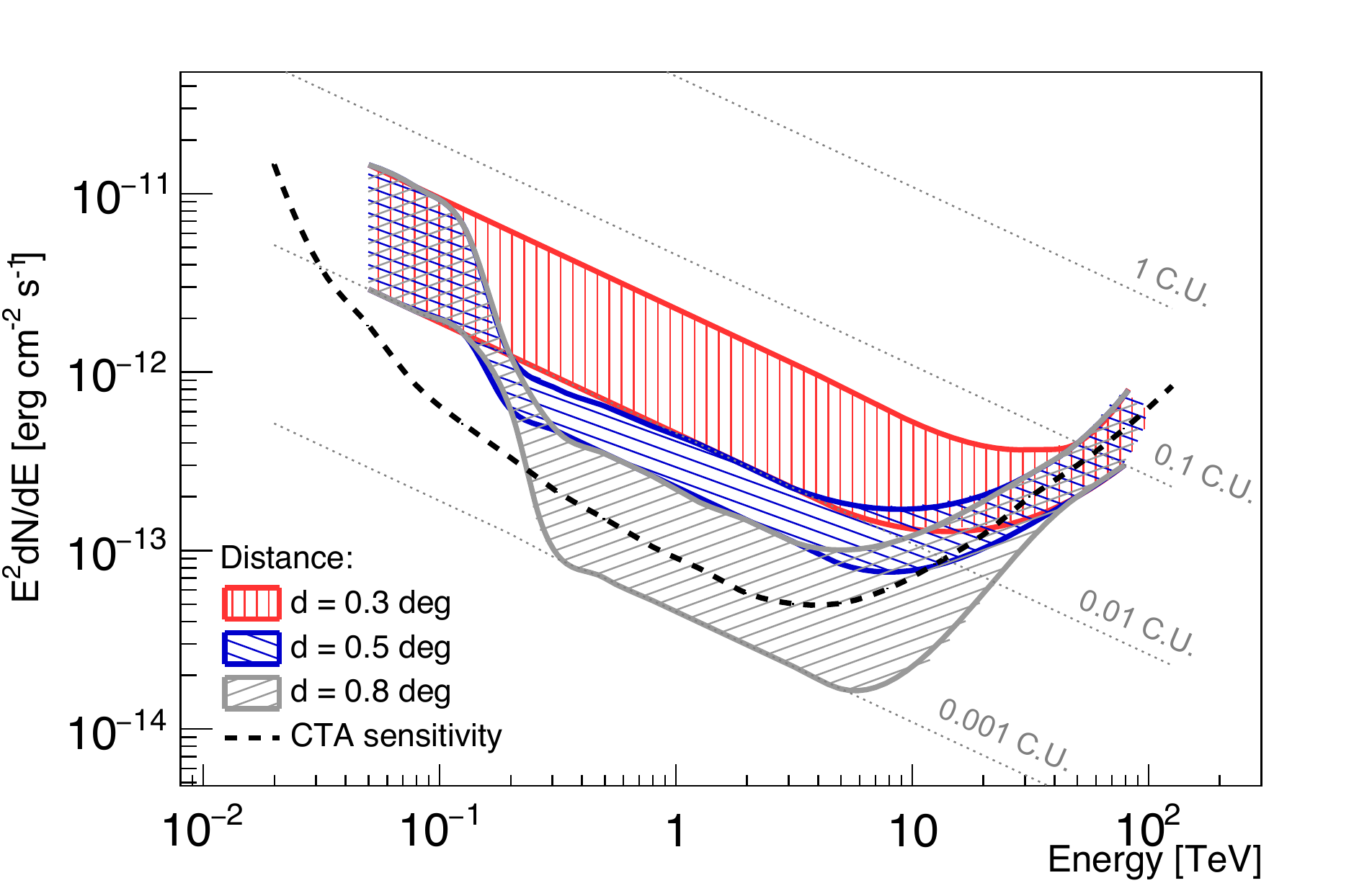}
\caption{Same as Fig. \ref{fig:sens13}, but for the case of non-Gaussian PSF with tails given by Eq. [\ref{eq:PSFtails}] with $K=0.3$.}
\label{fig:sens13_tails}
\end{figure}

The sensitivities shown in Figs. \ref{fig:sens11} and  \ref{fig:sens13} are obtained under the assumption that the PSF is described by a pure Gaussian distribution. However, in reality the PSF might be better represented by a broader distribution with long tails. In that case, the sensitivity around strong gamma-ray sources will be even more strongly reduced. The effect of the tails in the PSF on the sensitivity is shown in Figs. \ref{fig:sens11_tails} and \ref{fig:sens13_tails} (for point like and extended \emph{background source}, respectively). It is seen that for the assumed shape of the non-Gaussian PSF with tails (calculated for $K=0.3$) the zone of the reduced sensitivity extends up to $\sim 1$deg.

\section{Summary}

The sensitivity of the current systems of imaging atmospheric Cherenkov telescopes is limited by the background contributed by air showers induced by cosmic rays. It is expected that the next generation instruments, such as the CTA observatory, will significantly improve the sensitivity of the current instruments, like H.E.S.S., MAGIC and VERITAS. The larger effective area, better angular resolution, and more effective discrimination of hadronic showers will contribute to the reduction, by up to an order of magnitude, of the minimum detectable fluxes for point-like sources detected by CTA. However, the improvement of the sensitivity could be more modest if the \emph{test source} is located in complex environments densely populated by gamma-ray emitters, due to the additional background contributed by the gamma-rays from nearby sources. In this paper, we have quantitatively studied this effect under different assumptions regarding the shape of the PSF. It is demonstrated that for even a relatively modest background source of strength $\sim0.1$\,Crab, the background from a neighboring source can exceed the cosmic-ray background at distances up to $0.3$\,deg or even $\sim1$\,deg if the PSF is characterized by non-negligible tails. Depending on the PSF shape and the distance to the \emph{test source}, the minimum detectable flux is increased by a factor of a few or even by an order of magnitude. This implies an increase  by one to two orders of magnitude of the required minimum observation time. This should be taken into account when planning the observations with future imaging atmospheric Cherenkov arrays for specific objects, especially in the Galactic plane. Generally, this statement concerns all energy intervals. However, one should note that the situation could be more optimistic at multi-TeV energies. In fact, above $10$\,TeV (i) the energy resolution becomes better, (ii) the tails of the PSF are predicted to be more compact thanks to the higher telescope multiplicity per event (which would reduce the fluctuations responsible for the tails), and (iii) a (relatively) small number of multi-TeV sources (PeVatrons) is expected.

\section*{Acknowledgement}
The authors would like to thank T. Hassan and M. C. Macarone for their comments on the manuscript which improve greatly the paper. They also gratefully thank A. Taylor for carefully
checking the text. E. O. W. acknowledges the supports from the grants AYA2012-39303 and SGR2014-1073.


\bibliographystyle{elsarticle-num} 
\bibliography{bibliography}

\begin{thebibliography}{10}
\expandafter\ifx\csname url\endcsname\relax
  \def\url#1{\texttt{#1}}\fi
\expandafter\ifx\csname urlprefix\endcsname\relax\def\urlprefix{URL }\fi
\expandafter\ifx\csname href\endcsname\relax
  \def\href#1#2{#2} \def\path#1{#1}\fi

\bibitem{0034-4885-71-9-096901}
F.~Aharonian, J.~Buckley, T.~Kifune, G.~Sinnis,
  \href{http://stacks.iop.org/0034-4885/71/i=9/a=096901}{High energy
  astrophysics with ground-based gamma ray detectors}, Reports on Progress in
  Physics 71~(9) (2008) 096901.
\newline\urlprefix\url{http://stacks.iop.org/0034-4885/71/i=9/a=096901}

\bibitem{doi:10.1146/annurev.nucl.47.1.273}
F.~A. Aharonian, C.~W. Akerlof,
  \href{http://dx.doi.org/10.1146/annurev.nucl.47.1.273}{Gamma-ray astronomy
  with imaging atmospheric c粂erenkov telescopes}, Annual Review of Nuclear
  and Particle Science 47~(1) (1997) 273--314.
\newblock \href
  {http://arxiv.org/abs/http://dx.doi.org/10.1146/annurev.nucl.47.1.273}
  {\path{arXiv:http://dx.doi.org/10.1146/annurev.nucl.47.1.273}}, \href
  {http://dx.doi.org/10.1146/annurev.nucl.47.1.273}
  {\path{doi:10.1146/annurev.nucl.47.1.273}}.
\newline\urlprefix\url{http://dx.doi.org/10.1146/annurev.nucl.47.1.273}

\bibitem{Hinton:2009zz}
J.~A. Hinton, W.~Hofmann, {Teraelectronvolt astronomy}, Ann. Rev. Astron.
  Astrophys. 47 (2009) 523--565.
\newblock \href {http://arxiv.org/abs/1006.5210} {\path{arXiv:1006.5210}},
  \href {http://dx.doi.org/10.1146/annurev-astro-082708-101816}
  {\path{doi:10.1146/annurev-astro-082708-101816}}.

\bibitem{Hillas:2013txa}
A.~M. Hillas, {Evolution of ground-based gamma-ray astronomy from the early
  days to the Cherenkov Telescope Arrays}, Astropart. Phys. 43 (2013) 19--43.
\newblock \href {http://dx.doi.org/10.1016/j.astropartphys.2012.06.002}
  {\path{doi:10.1016/j.astropartphys.2012.06.002}}.

\bibitem{1742-6596-60-1-021}
D.~Horns, the HESS~Collaboration,
  \href{http://stacks.iop.org/1742-6596/60/i=1/a=021}{H.e.s.s.: Status and
  future plan}, Journal of Physics: Conference Series 60~(1) (2007) 119.
\newline\urlprefix\url{http://stacks.iop.org/1742-6596/60/i=1/a=021}

\bibitem{2008JPhCS.110f2009F}
R.~{Firpo}, {Status and results of the MAGIC telescope}, Journal of Physics
  Conference Series 110~(6) (2008) 062009.
\newblock \href {http://dx.doi.org/10.1088/1742-6596/110/6/062009}
  {\path{doi:10.1088/1742-6596/110/6/062009}}.

\bibitem{Holder:2008ux}
J.~Holder, et~al., {Status of the VERITAS Observatory}, AIP Conf. Proc. 1085
  (2009) 657--660.
\newblock \href {http://arxiv.org/abs/0810.0474} {\path{arXiv:0810.0474}},
  \href {http://dx.doi.org/10.1063/1.3076760} {\path{doi:10.1063/1.3076760}}.

\bibitem{Aharonian:2000rf}
F.~A. Aharonian, A.~K. Konopelko, H.~J. Volk, H.~Quintana, {A 5-GeV energy
  threshold array of imaging atmospheric Cherenkov telescopes at 5 km
  altitude}, Astropart. Phys. 15 (2001) 335--356.
\newblock \href {http://arxiv.org/abs/astro-ph/0006163}
  {\path{arXiv:astro-ph/0006163}}, \href
  {http://dx.doi.org/10.1016/S0927-6505(00)00164-X}
  {\path{doi:10.1016/S0927-6505(00)00164-X}}.

\bibitem{CTA}
M.~{Actis}, G.~{Agnetta}, F.~{Aharonian}, A.~{Akhperjanian}, J.~{Aleksi{\'c}},
  E.~{Aliu}, D.~{Allan}, I.~{Allekotte}, F.~{Antico}, L.~A. {Antonelli},
  et~al., {Design concepts for the Cherenkov Telescope Array CTA: an advanced
  facility for ground-based high-energy gamma-ray astronomy}, Experimental
  Astronomy 32 (2011) 193--316.
\newblock \href {http://arxiv.org/abs/1008.3703} {\path{arXiv:1008.3703}},
  \href {http://dx.doi.org/10.1007/s10686-011-9247-0}
  {\path{doi:10.1007/s10686-011-9247-0}}.

\bibitem{Hassan:2015bwa}
T.~Hassan, L.~Arrabito, K.~Bernlör, J.~Bregeon, J.~Hinton, T.~Jogler,
  G.~Maier, A.~Moralejo, F.~Di~Pierro, M.~Wood,
  \href{https://inspirehep.net/record/1389753/files/arXiv:1508.06075.pdf}{{Second
  large-scale Monte Carlo study for the Cherenkov Telescope Array}}, in:
  {Proceedings, 34th International Cosmic Ray Conference (ICRC 2015)}, 2015.
\newblock \href {http://arxiv.org/abs/1508.06075} {\path{arXiv:1508.06075}}.
\newline\urlprefix\url{https://inspirehep.net/record/1389753/files/arXiv:1508.06075.pdf}

\bibitem{Aharonian:2006pe}
F.~Aharonian, et~al., Observations of the crab nebula with h.e.s.s, Astron.
  Astrophys. 457 (2006) 899--915.
\newblock \href {http://arxiv.org/abs/astro-ph/0607333}
  {\path{arXiv:astro-ph/0607333}}, \href
  {http://dx.doi.org/10.1051/0004-6361:20065351}
  {\path{doi:10.1051/0004-6361:20065351}}.

\bibitem{Szanecki:2015zaa}
M.~Szanecki, D.~Sobczyńska, A.~Niedźwiecki, J.~Sitarek, W.~Bednarek, {Monte
  Carlo simulations of alternative sky observation modes with the Cherenkov
  Telescope Array}, Astropart. Phys. 67 (2015) 33--46.
\newblock \href {http://arxiv.org/abs/1501.02586} {\path{arXiv:1501.02586}},
  \href {http://dx.doi.org/10.1016/j.astropartphys.2015.01.008}
  {\path{doi:10.1016/j.astropartphys.2015.01.008}}.

\bibitem{Aharonian1997343}
F.~Aharonian, W.~Hofmann, A.~Konopelko, H.~Volk,
  \href{http://www.sciencedirect.com/science/article/pii/S0927650596000692}{The
  potential of ground based arrays of imaging atmospheric cherenkov telescopes.
  i. determination of shower parameters}, Astroparticle Physics 6~(324) (1997)
  343--368.
\newblock \href
  {http://dx.doi.org/http://dx.doi.org/10.1016/S0927-6505(96)00069-2}
  {\path{doi:http://dx.doi.org/10.1016/S0927-6505(96)00069-2}}.
\newline\urlprefix\url{http://www.sciencedirect.com/science/article/pii/S0927650596000692}

\bibitem{2013APh....43..171B}
K.~{Bernl{\"o}hr}, A.~{Barnacka}, Y.~{Becherini}, O.~{Blanch Bigas},
  E.~{Carmona}, P.~{Colin}, G.~{Decerprit}, F.~{Di Pierro}, F.~{Dubois},
  C.~{Farnier}, S.~{Funk}, G.~{Hermann}, J.~A. {Hinton}, T.~B. {Humensky},
  B.~{Kh{\'e}lifi}, T.~{Kihm}, N.~{Komin}, J.-P. {Lenain}, G.~{Maier},
  D.~{Mazin}, M.~C. {Medina}, A.~{Moralejo}, S.~J. {Nolan}, S.~{Ohm}, E.~{de
  O{\~n}a Wilhelmi}, R.~D. {Parsons}, M.~{Paz Arribas}, G.~{Pedaletti},
  S.~{Pita}, H.~{Prokoph}, C.~B. {Rulten}, U.~{Schwanke}, M.~{Shayduk},
  V.~{Stamatescu}, P.~{Vallania}, S.~{Vorobiov}, R.~{Wischnewski},
  T.~{Yoshikoshi}, A.~{Zech}, {CTA Consortium}, {Monte Carlo design studies for
  the Cherenkov Telescope Array}, Astroparticle Physics 43 (2013) 171--188.
\newblock \href {http://arxiv.org/abs/1210.3503} {\path{arXiv:1210.3503}},
  \href {http://dx.doi.org/10.1016/j.astropartphys.2012.10.002}
  {\path{doi:10.1016/j.astropartphys.2012.10.002}}.

\bibitem{Maier:2015bta}
G.~Maier, L.~Arrabito, K.~Bernlöhr, J.~Bregeon, F.~Di~Pierro, T.~Hassan,
  T.~Jogler, J.~Hinton, A.~Moralejo, M.~Wood,
  \href{https://inspirehep.net/record/1389675/files/arXiv:1508.06042.pdf}{{Monte
  Carlo Performance Studies of Candidate Sites for the Cherenkov Telescope
  Array}}, in: {Proceedings, 34th International Cosmic Ray Conference (ICRC
  2015)}, 2015.
\newblock \href {http://arxiv.org/abs/1508.06042} {\path{arXiv:1508.06042}}.
\newline\urlprefix\url{https://inspirehep.net/record/1389675/files/arXiv:1508.06042.pdf}

\bibitem{lima}
T.-P. Li, Y.-Q. Ma, {Analysis methods for results in gamma-ray astronomy}, apj
  272 (1983) 317--324.

\bibitem{Knodlseder:2014bka}
J.~Knödlseder, S.~Brau-Nogué, C.~Deil, C.-C. Lu, P.~Martin, M.~Mayer,
  A.~Schulz, {Cherenkov Telescope Array science data analysis using the
  ctools}, Proc. SPIE Int. Soc. Opt. Eng. 9152 (2014) 91522V.
\newblock \href {http://dx.doi.org/10.1117/12.2055210}
  {\path{doi:10.1117/12.2055210}}.

\bibitem{Becherini:2012iy}
Y.~Becherini, B.~Khelifi, S.~Pita, M.~Punch, {Advanced analysis and event
  reconstruction for the CTA Observatory}, AIP Conf. Proc. 1505 (2012)
  769--772.
\newblock \href {http://arxiv.org/abs/1211.5997} {\path{arXiv:1211.5997}},
  \href {http://dx.doi.org/10.1063/1.4772373} {\path{doi:10.1063/1.4772373}}.

\bibitem{Aharonian:2004gb}
F.~Aharonian, et~al., {The Crab nebula and pulsar between 500-GeV and 80-TeV.
  Observations with the HEGRA stereoscopic air Cerenkov telescopes}, Astrophys.
  J. 614 (2004) 897--913.
\newblock \href {http://arxiv.org/abs/astro-ph/0407118}
  {\path{arXiv:astro-ph/0407118}}, \href {http://dx.doi.org/10.1086/423931}
  {\path{doi:10.1086/423931}}.

\bibitem{Carrigan:2013}
S.~Carrigan, the HESS~Collaboration, {The H.E.S.S. Galactic Plane Survey -
  maps, source catalog and source population}, in: {Proceedings, 33rd
  International Cosmic Ray Conference (ICRC 2015)}, 2013.
\newblock \href {http://arxiv.org/abs/1307.4690} {\path{arXiv:1307.4690}}.

\end{thebibliography}

\end{document}